\documentclass[nofootinbib,twocolumn,aps,pre,superscriptaddress,citeautoscript,floatfix]{revtex4-1}
\usepackage{graphicx}
\usepackage{amsmath,amssymb}
\usepackage[caption=false]{subfig}

\usepackage[dvipsnames]{xcolor}

\begin{document}
\title{Charge Regulating Macro-ions in Salt Solutions:\\
Screening Properties and Electrostatic Interactions
}
\author{Yael Avni}
\affiliation{Raymond and Beverly Sackler School of Physics and Astronomy,\\ Tel Aviv University, Ramat Aviv 69978, Tel Aviv, Israel}
\author{Tomer Markovich}
\affiliation{Raymond and Beverly Sackler School of Physics and Astronomy,\\ Tel Aviv University, Ramat Aviv 69978, Tel Aviv, Israel}
\affiliation{DAMTP, Centre for Mathematical Sciences, University of Cambridge, Cambridge CB3 0WA, United Kingdom}
\author{Rudolf Podgornik}\thanks{Also affiliated with the Department of Physics, Faculty of Mathematics and Physics, University of Ljubljana, and with the
Department of Theoretical Physics, J. Stefan Institute, 1000 Ljubljana, Slovenia}
\affiliation{School of Physical Sciences and Kavli Institute for Theoretical Sciences, University of Chinese Academy of Sciences, Beijing 100049, China
and \\CAS Key Laboratory of Soft Matter Physics, Institute of Physics, Chinese Academy of Sciences, Beijing 100190, China}
\author{David Andelman}
\email{andelman@post.tau.ac.il}
\affiliation{Raymond and Beverly Sackler School of Physics and Astronomy,\\ Tel Aviv University, Ramat Aviv 69978, Tel Aviv, Israel}


\begin{abstract}
We revisit the charge-regulation mechanism of macro-ions and apply it to mobile macro-ions in a bathing salt solution. In particular, we examine the effects of correlation between various adsorption/desorption sites and analyze the collective behavior in terms of the solution effective screening properties. We show that such a behavior can be quantified in terms of the charge {\em asymmetry} of the macro-ions, defined by their preference for a non-zero effective charge, and their {\em donor/acceptor} propensity for exchanging salt ions with the bathing solution. Asymmetric macro-ions tend to increase the screening, while symmetric macro-ions can in some cases decrease it. Macro-ions that are classified as donors display a rather regular behavior, while  those that behave as acceptors exhibit an anomalous non-monotonic Debye length.
The screening properties, in their turn, engender important modifications to the disjoining pressure between two charged surfaces. Our findings are in particular relevant for solutions of proteins, whose exposed amino acids can undergo charge dissociation/association processes to/from the bathing solution, and can be considered as a solution of charged regulated macro-ions, as analyzed here.
\end{abstract}

\maketitle

\section{Introduction}
Long-range interactions between biological macromolecules are similar in many respects to those that characterize inorganic colloids. Following the Deryaguin-Landau-Verwey-Overbeek (DLVO) paradigm~\cite{Safinya}, the interactions can be decomposed into the van der Waals and the electrostatic components~\cite{Woods2016,Israelachvili1991}. However, more complex examples of colloids with {\em ionizable groups}, or proteins with ionizable amino-acid residues~\cite{Perutz1978, Warshel2006, Gitlin2006} differ substantially from colloids that carry a fixed charge~\cite{Verwey1948}. Dissociation of such chargeable moieties engenders an exchange of ions (usually but not necessarily, a proton, H$^{+}$) with the bathing solution~\cite{Borkovec1,Borkovec2}.  Consequently, it changes the nature of protein-protein interactions~\cite{Lund2005,Fer2009}, and modifies the protein-specific spatial charge distribution~\cite{Tanford1967, Warshel2006, Anze}.

The exchange of ions between proteins (via their dissociable groups) and the surrounding solution has been  addressed already  in the 1920's by Linderstr{\o}m-Lang of the Carlsberg Laboratory~\cite{Linderstrom-Lang}. Later on, it was referred to as the {\em charge regulation} (CR) mechanism, and its formalism was set fourth by Ninham and Parsegian~\cite{NP-regulation}. In their seminal 1971 work, CR was formulated within the Poisson-Boltzmann (PB) theory of electrostatic interactions in an aqueous environment~\cite{Safinya}, which included an additional self-consistent boundary condition at the CR bounding surfaces. In recent decades, the CR formulation was implemented for surface binding sites via the law of mass action~\cite{Regulation2, Regulation3, Regulation4, Regulation5}, and separately, by modifying the surface free-energy~\cite{CTAB, Pincus, Olvera, Olvera2, Natasa1, Maggs, epl1, Markovich2016EPL, Diamant, Maarten,Ben-Yaakov2, Biesheuvel2005}. The latter approach leads to the same surface dissociation equilibrium as does the law of mass action, but with the advantage that it can be easily generalized to include other non-electrostatic surface interactions.

\begin{figure*}
\includegraphics[width = 1.8\columnwidth,draft=false]{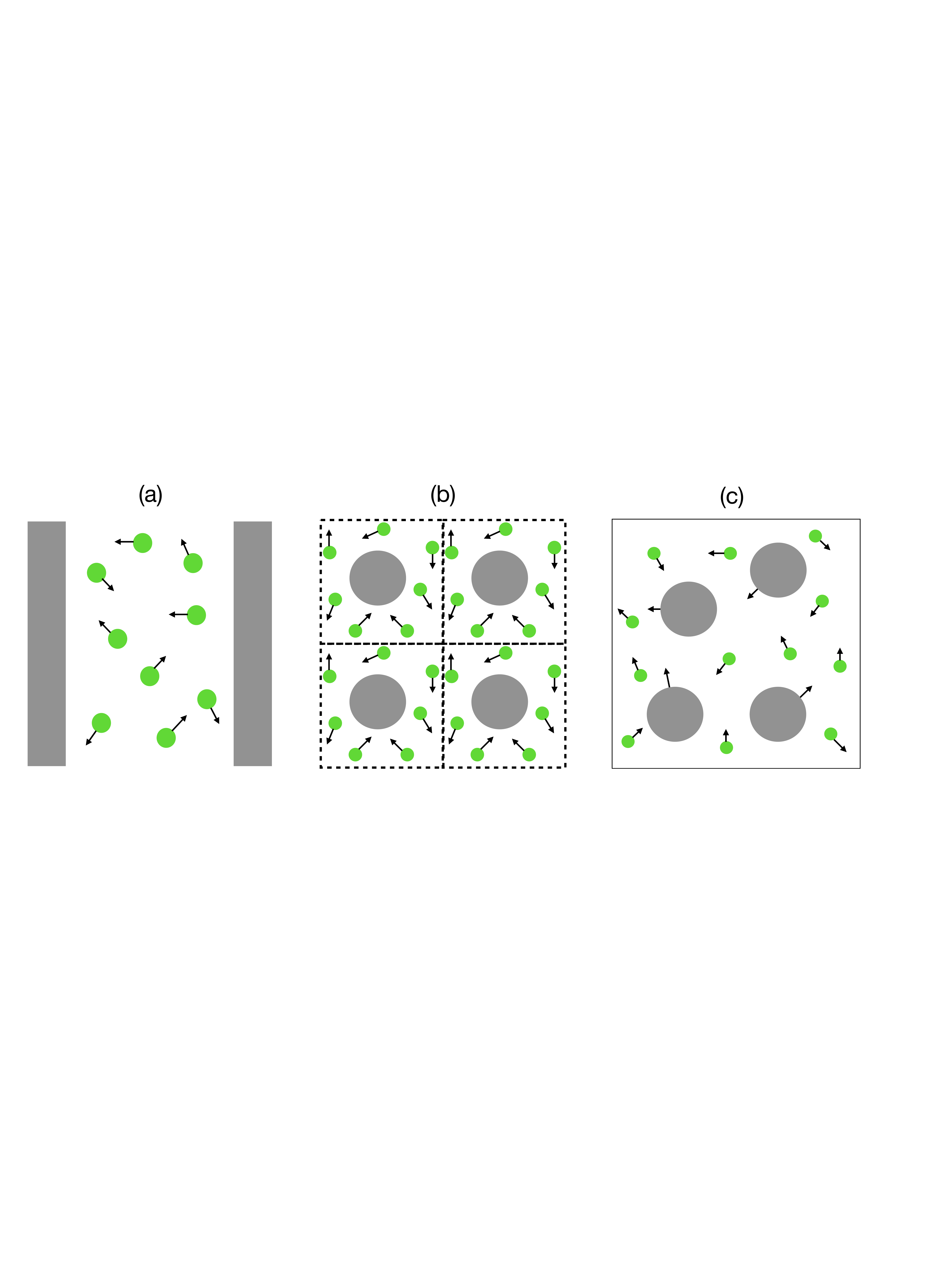}
\caption{\textsf{(color online)   A schematic illustration of different approaches to model charge regulating macro-ions (gray) in an ionic solution (green). (a) The macro-ions are approximated by flat  bounding surfaces, with salt ions moving between them. (b) Each macro-ion is fixed in a cell with mobile salt ions within the cell. The interaction between the neighboring macro-ions is mimicked by the boundary conditions of the cell. (c) Both the salt ions and  macro-ions are free to move in the entire system. While (a) and (b) were modeled and analyzed extensively in the past, our theory strives to describe the latter case (c), which is the only one that accounts for the collective behavior of the macro-ions.}}
\label{Fig1}
\end{figure*}

The charge association/dissociation process (CR mechanism) couples the local electrostatic field with the local charge, and results in a partition of dissociated and associated states,  which is obtained self-consistently~\cite{Markovich2016EPL}. This local coupling implies a complex dependence of the net charge of the macro-ion, protein or proteinaceous aggregate on the solution conditions, the local dielectric profiles and the overall system geometry~\cite{Nap}. Furthermore, in some cases, higher-order electrostatic multipoles may need to be considered in relation to the CR process, in addition to its monopolar ones~\cite{Anze}.

The PB theory with CR charges was  studied in the past for planar geometries of one or two charged-regulating surfaces, in contact with a bathing electrolyte solution, or by modeling a charged colloid in solution, in the proximity of another charged surface~\cite{ninham1971electrostatic, chan1976electrical, Markovich2016EPL, PhysRevLett.117.098002}. Other realistic geometries of protein-protein interactions in different aqueous solution environments have also been  studied by various simulation techniques~\cite{Lund2,Lund1}.

A viable simplification of the above problem is described by a variant of the {\em cell model}~\cite{Alexander1984, borkovec2001ionization, borkovec1997difference, ullner1994conformational, boon2012charge}, where the macro-ion is enclosed within an external cell, whose boundary conditions mimic the presence of neighboring macro-ions. We note that the cell model, as well as the other CR works mentioned above, neglects the translational entropy of the macro-ions, and can be justified only when the salt concentration is high enough, such that collective many-body effects become irrelevant.
In the opposite limit, when the translational degrees of freedom of macro-ions and salt ions are strongly coupled, one needs to employ a more refined and collective description.

To this effect, a model based on the PB framework has been recently developed by us~\cite{Markovich2017EPL} to account for collective CR effects of {\em mobile} macro-ions in dilute salt solutions. The model deals with charge regulating macro-ions and simple salt ions on an equal footing (see Fig.~\ref{Fig1}). As an example, it was implemented so far only for a single, specific CR model of the macro-ions.

Motivated by the recent studies of charge regulation of nonpolar colloids~\cite{Bartlett}, we extend our  approach~\cite{Markovich2017EPL} by studying in detail different generalized CR models of mobile macro-ions in electrolyte solutions. The main focus is on {\em zwitterionic} macro-ions, {\em i.e.},
dissociable macro-ions that contain both positive and negative charge states. For example, proteins with protonation/deprotonation of chargeable solvent exposed amino acids. In such cases, the more subtle coupling between the charge regulating macro-ions and fully dissociated salt ions leads to completely unexpected and important modifications.

Hereafter, we present a few intriguing examples where the effective screening length exhibits a {\em non-monotonic } and steep dependence both on the salt and charge regulating macro-ion bulk concentrations. The mechanism responsible for this behavior is unraveled in terms of the charge asymmetry of the macro-ions and their {\em donor/acceptor} propensity. Furthermore, the disjoining pressure between two charged surfaces is shown to exhibit a non-monotonic dependence on the macro-ion bulk concentration at a fixed separation between the surfaces. These unexpected features of
dissociating macro-ions shed new light on the understanding of electrostatic interactions in complex colloid systems.

The outline of our paper is as follows. In Sec.~\ref{model}, we present
a general approach of treating a solution containing mobile CR macro-ions and its free energy. We then introduce, in Sec.~\ref{results}, two generic types of CR mechanisms for zwitterionic macro-ions, and obtain the corresponding mean-field description of their electrostatic interactions. We further calculate  the modified screening length and disjoining pressure as applied to two charged planar surfaces immersed in solutions containing mobile CR macro-ions. In Sec.~\ref{discussion}, we conclude with several more general observations and remarks about connection to existing experiments and suggest future ones.

\section{The Charge-Regulation Model} \label{model}
Our model system is composed of monovalent salt ions in an aqueous solvent with dielectric permittivity $\varepsilon$ at temperature $T$. Macro-ions are added to the solution in the form of colloidal particles with active charge groups, enabling the desorption/adsorption of monovalent ions to/from the solution. For simplicity, the dissociating ions are assumed to be of the same type as the salt ions, so there are only two types of small ions in solution: monovalent cations and anions. This assumption simplifies the calculation but can be easily relaxed by rather minor modifications to the model.

Before macro-ions are added to the aqueous solution, they are assumed to be neutral. As they are put in contact with the bathing solution, they become charged because of the dissociation process. It is convenient to define the bulk concentration of the monovalent salt (prior to the addition of macro-ions) as $n_0$, and the bulk concentration of the added macro-ions as $p_{b}$, because these are the two experimentally controllable parameters. Note that after the macro-ions are added and equilibrate with the ionic solutes, the bulk cation and anion  concentrations, will differ from $n_0$. We define them as $n^{+}_b$ and $n^{-}_b$, and they depend on $n_0$ as well as on $p_b$. We will see that this dependence hinges on the details of the CR model, and is different for the two specific CR models considered below.

For clarity sake, we assume that the solvent molecules and salt ions have the same volume, $a^3$, whereas the macro-ion specific volume is written as $\gamma a^3$, where $\gamma>1$ is a numerical pre-factor describing the ratio between the two molecular volumes. As the effective radius of a protein is typically $1-5$\,nm for molecular weights in the range of $5-500$\,kDa~\cite{Erickson}, while the simple salt ions have a typical size of $\sim 0.3$\,nm, the corresponding $\gamma$ has values up to $\sim 10^{3}$. Much higher values of $\gamma$ would have to be considered for CR nano-particles, entailing a consistent inclusion of packing effects at higher concentrations, a direction that we will not pursue further in this work.


\begin{figure*}
\includegraphics[width = 1.8\columnwidth,draft=false]{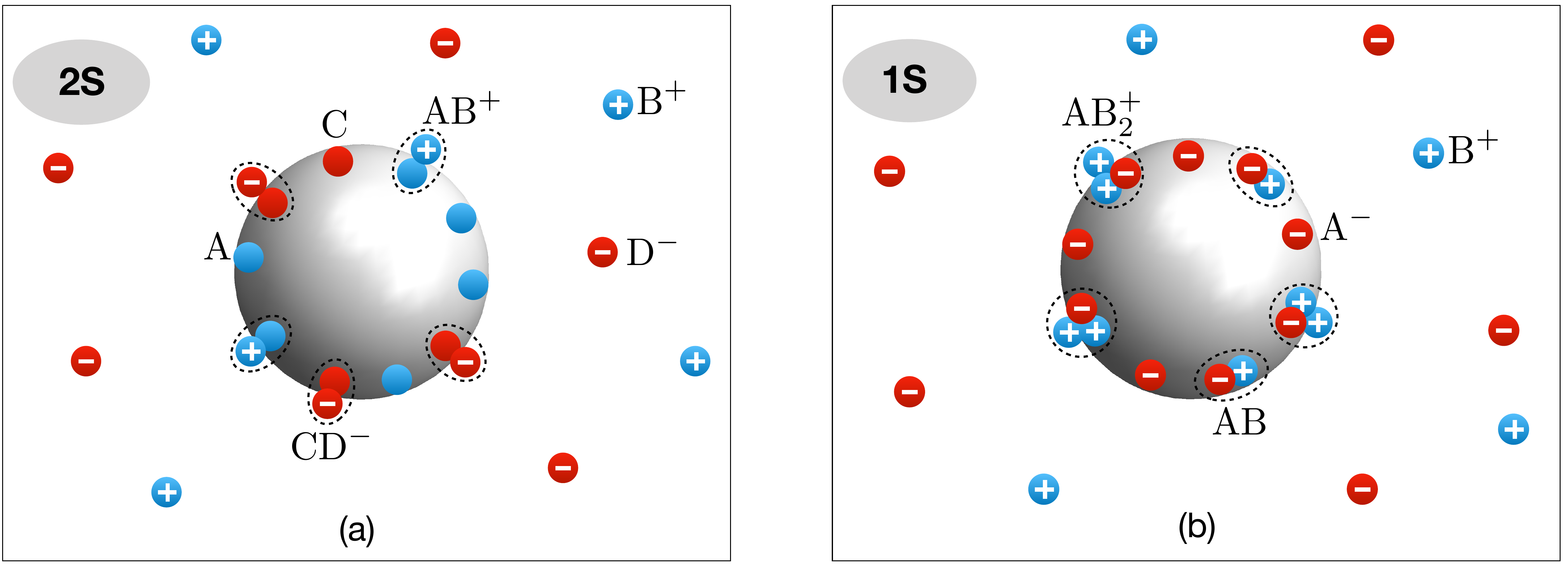}
\caption{\textsf{(color online)  The two charge regulation models, 2S and 1S, described by Eqs.~(\ref{reaction2S}) and (\ref{reaction1S}), respectively.
(a) In the 2S model, each macro-ion has two types of surface dissociable sites, A and C. The A site can adsorb/desorb a cation ($\rm B^+$), while the C site can adsorb/desorb an anion (D$^{-}$). During the adsorption process, the neutral A site acquires a positive cation, becoming AB$^+$, while the neutral C site becomes negatively charged as CD$^{-}$.
(b) In the 1S model, only cations can be adsorbed from solution. The macro-ion carries only one type of active surface sites, and each such  site has three charging states. (i)~It can stay neutral in its native AB state;  (ii)~it can release a cation ($\rm B^+$) and become negative ($\rm A^-$); or,  (iii)~ it can adsorb a cation ($\rm B^+$) and become positive (${\rm AB_2^{+}}$).
}}
\label{Fig2}
\end{figure*}
In the dilute limit of both solutes (salt and macro-ions), the entropy can be approximated by the ideal  entropy of mixing
\begin{equation}
\label{s1}
S/k_{B} = -\sum_{i=\pm}n_{i}\left[\ln(n_{i}a^{3})-1\right]
\,\,-\,\,
p\left[\ln(p\gamma a^{3})-1\right],
\end{equation}
while the total mean-field free energy, $F=U-TS$ has the form~\cite{Markovich2017EPL,Safinya,Maggs2016general}:
\begin{eqnarray}
\label{bp1}
 F &=& \int_{\rm V}\text{d}^{3}r \Bigg[ -\frac{\varepsilon_0 \varepsilon}{2}\left(\nabla\psi\right)^{2} +e(n_{+}-n_{-})\psi \,\, -TS \nonumber\\
 &+& pg(\psi)   - (\mu_{+}n_{+}  +  \mu_{-}n_{-}  +  \mu_{p} p ) \Bigg] \, ,
\end{eqnarray}
where $n_\pm({\bf r})$ and $p({\bf r})$ are the number densities at position ${\bf r}$ of the small $\pm$ ions and  macro-ions, respectively, $k_{\mathrm{B}}$ is the Boltzmann constant, $\psi({\bf r})$ is the local electrostatic potential, $\mu_\pm $ are the chemical potentials of the monovalent ions and $\mu_{p}$ for the macro-ions.
The macro-ion term in the above expression is given by the single macro-ion free energy, $g(\psi )$, that characterizes the specific CR process. The functional dependence of $g(\psi )$ on other system parameters differs for the two CR models that we consider. In fact, this is the only non-standard term of Eq.~(\ref{bp1}), first introduced in Ref.~\cite{Markovich2017EPL} and shown to have significant consequences.

As the CR process takes place on the surface of the macro-ion located at position $\bf{r}$, the corresponding $g(\bf{r})$ can be written as
\begin{equation}
g \left( \bf{r} \right) = \oint _{\rm S} {\rm d}^2{\bf r'} g_{\rm s} \left(\bf{r} +\bf {r'}\right),
\label{lim1}
\end{equation}
where the form of the surface free-energy $g_{\rm s}(\bf r)$ is identical to that used in the CR models formulated for dissociable surfaces, and can assume various forms including Langmuir-Davies and Langmuir-Frumkin-Davies isotherms~\cite{Markovich2016EPL,Harries,Ben-Yaakov2,Safinya}.
We exploit the fact that as long as the inter-particle typical distances are larger than the macro-ions size ($p \gamma a^3 \ll 1$, $n_{\pm} \gamma a^3 \ll 1$),
$g_{\rm s}\left({\bf{r}}\right)$ can be treated as a purely local function at position ${\bf r}$.
When the macro-ions are densely packed and/or highly charged, the point-particle mean-field theory breaks down. In that case one needs to include the finite size of the macro-ions and the non-uniform distribution of sites on the surface, along with correlations beyond the mean-field approximation.

The thermodynamical equilibrium is obtained by using the variational principle for $F$ with respect to all the thermally annealed variables: the three densities $n_\pm, p$, the potential $\psi$, and the fraction of associated ions, to be introduced in the next section. The variation with respect to the three densities and electrostatic potential can be taken without any prior knowledge of $g(\psi)$.

From $\delta F/\delta n_{\pm}=0$, it follows that the ion densities $n_\pm$ satisfy the Boltzmann distribution,
\begin{equation}
\label{n1}
n_{\pm}(\psi)=n^\pm_{b}{\rm e}^{\mp \beta e\psi},
\end{equation}
where $\beta=1/k_B T$ and $n^{\pm}_{b}= \exp(\beta\mu_\pm)/a^3$ is defined as the cation/anion bulk concentration, taken at zero reference potential, $\psi=0$.

Next, the variation, $\delta F/\delta p=0$, gives the Boltzmann distribution for the macro-ion concentration,
\begin{equation}
\label{p}
p(\psi) = p_b {\rm e}^{-\beta g(\psi) + \beta g_0} \, ,
\end{equation}
where $g_0=g(\psi=0)$ acts as a reference bulk state with $\psi=0$, and the macro-ion bulk density of the macro-ion $p_b$ satisfies
$p_b = \exp( \beta \mu_p -\beta  g_0)/\gamma a^3$\,.

The variation of $F$ with respect to $\psi$ finally yields a generalized Poisson-Boltzmann equation,
\begin{eqnarray}
\label{PB}
-\varepsilon_0 \varepsilon\nabla^2\psi = \rho(\psi) = e [n_{+}(\psi) - n_{-}(\psi)] + Q(\psi) p(\psi) \, ,
\end{eqnarray}
where the effective charge of each macro-ion is defined as
\begin{eqnarray}
\label{ep}
Q = \frac{\partial g(\psi)}{\partial\psi} \, .
\end{eqnarray}
Note that if the macro-ions were simple ions of valency $\pm z$, the effective charge would be $Q=\pm ze$. Consequently, $g$ would have the usual form $\pm e z \psi$, and Eqs.~(\ref{p})-(\ref{ep}) would reduce to the standard form of the Poisson-Boltzmann equation.

\section{Results and Discussion}\label{results}

We present our results for two different CR models, each with a distinct macro-ion free energy, $g(\psi)$.
The variation of the free energy  as in Eqs.~(\ref{n1})-(\ref{ep}) together with the variation with respect to the CR degrees of freedom on the macro-ions, leads to a complete set of equations of state for the two CR models. By solving them, we can get insightful predictions about the screening length and the disjoining pressure between two charged surfaces.

\subsection{The two-site (2S) model} \label{2S model}

We consider as our first CR model the case where the zwitterionic  macro-ions contain two types of active sites:
one that can adsorb/desorb a cation from/to the solution, and the other site that can adsorb/desorb an anion (see Fig.~\ref{Fig2}(a)).
This adsorption model, involving two distinct types of active sites,  is denoted hereafter as the {\em 2S model}.
It is described by the chemical reaction equations:
\begin{equation}
\label{reaction2S}
\begin{split}
\text{A}+\text{B}^{+} & \rightleftarrows \text{A}\text{B}^{+}\\
\text{C}+\text{D}^{-} & \rightleftarrows \text{C}\text{D}^{-},
\end{split}
\end{equation}
where $\text{A}$ and $\text{C}$ represent two active macro-ion sites. The charges, $\text{B}^{+}$ and $\text{D}^{-}$ are, respectively, monovalent cation and anion, which can be released to the solution or bind onto the macro-ion from the solution (see Fig.~\ref{Fig2}(a)).
Such a model was considered in Ref.~\cite{Everts2016} to explain the melting of electrostatically repelling colloids upon increasing the colloid density, and was studied experimentally in Ref.~\cite{Bartlett}.

We assume that each macro-ion contains $N_{+}$ potentially dissociable $\text{A}$ sites and $N_{-}$ dissociable $\text{C}$ sites. Note that only a fraction of the overall $N_{\pm}$ sites will become charged at any given solution condition. This fraction is introduced below as $\phi_\pm$. Furthermore, the A and C active sites are uncorrelated in any direct manner.

In the 2S model, the free energy of a single macro-ion, $g(\psi)$, is assumed to be given by
\begin{equation} \label{g2S}
\begin{split}
g^{\rm 2S}(\psi,\phi_{+},\phi_{-}) & =
\sum_{i=\pm} \, N_{i}\phi_{i}(e_{i}\psi-\mu_{i}-\alpha_{i})\\
  +\,k_{B}T\sum_{i=\pm} & N_{i}\Bigl[\phi_{i}\ln \phi_{i} +(1-\phi_{i})\ln(1-\phi_{i})\Bigr] \, ,
\end{split}
\end{equation}
where $e_{\pm}=\pm e$, $\phi_{+}$   ($\phi_{-}$) is the fraction of the total $N_{+}$   ($N_{-}$) sites
that adsorb a cation (anion) from solution, and $\alpha_{\pm}$ is the free-energy change in adsorbing a cation/anion, respectively.
If $\alpha_\pm>0$, it means that there  is a free-energy gain of association, while a negative $\alpha_\pm$ opposes such binding.\footnote{Note that the model analyzed in Ref.~\cite{Markovich2017EPL} is a special case of the 2S model. It can be obtained in the following special limit:
$N_{+}=2N_{-}$ and $\alpha_{-}\to \infty$, so that $\phi_{-}=1$ for negative charges.}
It is possible to consider more sophisticated forms of $g(\psi)$, for example by including an interaction term between the different sites. In this paper we neglect such interactions, and justify it by assuming that the different sites in the macro-ion are situated far apart and interact only weakly.

We can now take the  variation of $F$ with respect to the occupied ionic fractions $\phi_\pm$, $\delta F/\delta\phi_\pm=0$. This gives the Langmuir-Davies isotherm~\cite{Diamant,Harries}
\begin{equation} \label{phi2S}
\phi_{\pm}=\frac{{Z}^{\rm 2S}_{\pm}(\psi)-1}{Z^{\rm 2S}_{\pm}(\psi)},
\end{equation}
where
\begin{equation} \label{Z2S}
Z^{\rm 2S}_{\pm}
= 1+n_b ^{\pm}K_\pm \, {\rm e}^{ \mp  \beta e \psi},
\end{equation}
is the 2S partition function of a single macro-ion, and throughout the remaining of this subsection we shall omit the 2S superscript.
In the above equation, we express $Z_{\pm}$ in terms of the chemical equilibrium constants, $K_{\pm}$~\cite{Safinya,healy1978ionizable},
\begin{equation}\label{kinC}
  K_\pm=a^{3}{\rm e}^{\beta\alpha_\pm} .
\end{equation}
Note that each of the $K_\pm$ varies between zero and infinity corresponding to $\alpha_\pm$ varying between $-\infty$ and $\infty$.
The meaning of the limit $K_\pm\to \infty$ (or $\alpha_\pm\to \infty$) is that all active sites are fully associated and charged in the bulk ($\phi_\pm\to 1$).
Since in experiments only $K_{\pm}$ is the measurable quantity, and not  $a$ or $\alpha_{\pm}$ separately, we will use hereafter
$K_\pm$ as the natural parameter in the numerical calculations and figures\footnote{If the dissociating ion is a proton, the relation between the pH, pK and our parameters is: $K_+ n_+=10^{\rm{pK-pH}}$.}.

Substituting the expression for $\phi_\pm$, Eq.~(\ref{phi2S}), into Eq.~(\ref{g2S}), we obtain
\begin{equation}
\label{g2S2}
g(\psi)=-k_{B}T\sum_{i=\pm}N_{i}\ln\left(Z_{i}\right) \, .
\end{equation}
Using Eqs.~(\ref{p}), (\ref{ep}) and (\ref{g2S2}), we write the macro-ion concentration $p(\psi)$  as
\begin{equation} \label{p2S}
p(\psi)=
p_{b}\left(\frac{Z_{+}(\psi)}{Z_{+}(0)}\right)^{N_{+}} \left(\frac{Z_{-}(\psi)}{Z_{-}(0)}\right)^{N_{-}} ,
\end{equation}
and its total charge, $Q(\psi)$
\begin{eqnarray}
\label{ep2S}
Q(\psi) & = & Q_{+}+Q_{-}=e\left(N_{+}\phi_{+}-N_{-}\phi_{-}\right) \nonumber\\
& = & e\left(N_{+}\frac{Z_{+}(\psi)-1}{Z_{+}(\psi)}-N_{-}\frac{Z_{-}(\psi)-1}{Z_{-}(\psi)}\right) \, .
\end{eqnarray}
Combining Eqs.~(\ref{p2S}), (\ref{ep2S}) and (\ref{n1}), the local charge density of cations, anions and charged macro-ions can be written explicitly in terms of $\psi$,
\begin{eqnarray}
\label{rho2S}
 \rho(\psi) &=& e\Big(  n^{+}_{b}{\rm e}^{-\beta e\psi} - n^{-}_{b}{\rm e}^{\beta e\psi}\Big)
+ p(\psi) Q(\psi),
\end{eqnarray}
with $p(\psi)$ and $Q(\psi)$ given in Eqs.~(\ref{p2S}) and (\ref{ep2S}).

Finally, we express the bulk concentrations, $n^\pm_{b}$, of the two ionic species in terms of the two controllable parameters: the bulk salt concentration before the addition of the macro-ions, $n_0$, and the bulk concentration of the macro-ions, $p_{b}$. Throughout most of the analysis we assume that the macro-ions are added to the solution when all of their sites are initially neutral, {\it i.e}, A and C are not associated to B$^+$ and D$^-$ (see Fig.~\ref{Fig2}(a)). In that case,

\begin{eqnarray}
\label{n_02S}
n^\pm_{b} = n_{0} - p_{b} N_{\pm}\phi_\pm(0).
\end{eqnarray}
The above form ensures that electro-neutrality is satisfied in the bulk, and is in agreement with the definition of $n_b^\pm$
and $n_0$ presented in the beginning of Sec.~II. Note that $\phi_\pm(0)$ also depends on $n_{b}^\pm$
via Eq.~(\ref{phi2S}). Thus,  Eq.~(\ref{n_02S}) is quadratic  in $n_b^\pm$, but it has only one physically admissible root.

\subsection{Screening length for 2S model} \label{Screening section}
After deriving the expressions for $n_\pm(\psi)$, $Q(\psi)$ and $p(\psi)$  in terms of the model parameters, we can analyze the effect of adding macro-ions and simple monovalent salt. A convenient way to proceed is by analyzing the effective screening length, $\lambda_{\rm eff}$. It is obtained by expanding the generalized PB equation, Eq.~(\ref{PB}), to the first order in the electrostatic potential, $\psi$,
\begin{eqnarray}
\label{PB1}
-\varepsilon_0 \varepsilon\nabla^2\psi = \rho(\psi) \simeq  \rho(0)+\rho'(0)\psi+\mathcal{O}\left(\psi^2\right) \, ,
\end{eqnarray}
and taking into account electroneutrality in the bulk, $\rho(0)=0$. The equation that follows is equivalent to the standard Debye-H\"uckel equation, except that the role of the Debye screening length, $\lambda_{\rm D}$, is played by an {\em effective screening length}, $\lambda_{\rm eff}$,
\begin{equation}
\label{screening}
\lambda_{\rm eff}^{-2} = -\frac{1}{\varepsilon_0 \varepsilon} \frac{\partial\rho}{\partial\psi} \Big|_{\psi=0} \,,
\end{equation}
where in the absence of macro-ions, the bulk salt concentration is $n_0$, and $\lambda_{\rm eff}$ reduces to the Debye screening length,
$\lambda_{\rm D}=1/\sqrt{2 e^2 n_{0} /(\varepsilon_0 \varepsilon {k_{\mathrm{B}}T} )}$.

To get the screening length in the 2S model, we substitute Eq.~(\ref{rho2S}) into the above equation and take into account the electroneutrality condition, Eq.~(\ref{n_02S}), wherefrom it follows that
\begin{equation}
\label{screening2S}
\lambda_{\rm eff}=  \lambda_{\rm D} \Bigg( 1 +
\frac{1}{2} \frac{p_{b}}{ n_0} \frac{Q^2(0)}{e^2}
- \frac{1}{2} \frac{p_{b}}{n_0} \Big[N_{+}\phi_{+}^{2}(0)+N_{-}\phi_{-}^{2}(0) \Big] \vphantom{\frac{A}{B}} \Bigg)^{-1/2} \, ,
\end{equation}
with $\phi_{\pm}$ and $Q$ given by Eqs.~(\ref{phi2S}) and (\ref{ep2S}), respectively. The deviation of $\lambda_{\rm eff}$ from $\lambda_{\rm D}$ is manifested in the two terms proportional to $p_b$, which depend on the number fractions $\phi_\pm$ and represent two opposite trends. The first of them, proportional to $Q^2(0)$, is positive and always decreases the screening length. The term in the square brackets is negative and increases the screening length. Nevertheless, it can be shown that the whole expression for $\lambda_{\rm eff}$ is always positive, irrespective of  parameter values.

\subsubsection{Symmetric vs. asymmetric macro-ions}
From Eq.~(\ref{screening2S}) one can see that macro-ions with $N_{+} \phi_{+} \approx N_{-} \phi_{-}$, {\em i.e.,} with an overall vanishing charge, tend to increase the screening length. Such macro-ions bind cations and anions on average {\em in pairs}, thus diminishing the effective salt concentration and increasing the screening length. We refer to such macro-ions as ``{\it symmetrically}" charged (or simply ``symmetric"). Note that the simplest case of symmetrically charged macro-ions is obtained for $K_{+} \simeq K_{-}$ and $N_{+} \simeq N_{-}$.

In the other case, ``{\it asymmetrically}" charged macro-ions, $N_{-} \phi_{-} \ll N_{+} \phi_{+}$ (or $N_{-} \phi_{-} \gg N_{+} \phi_{+}$), decrease the screening length. As asymmetric macro-ions bind preferably either cations or anions but not both, the charged sites themselves start acting as screening agents, leading to a decrease in screening length.

Charge asymmetry of macro-ions appears as an important feature determining the effective screening. We note that for charged polymers, the asymmetric case corresponds to net charged polyelectrolytes, while the symmetric one corresponds to polyampholytes, containing positive and negative charged groups, which approximately balance each other~\cite{Borukhov}.
%

\begin{figure}[h!]
\includegraphics[width=1\columnwidth,draft=false]{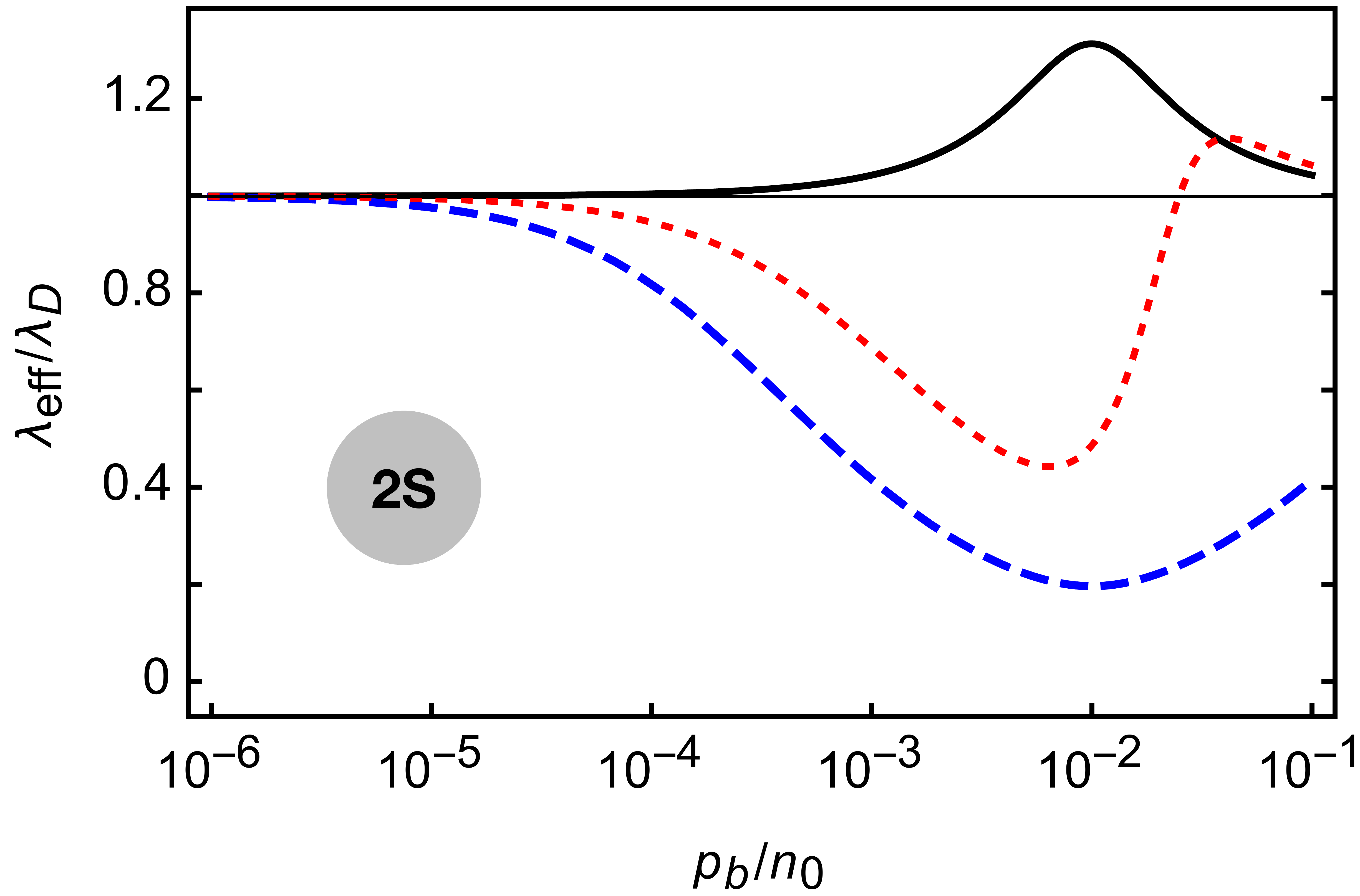}\centering
\caption{\textsf{(color online) The screening length in the 2S model, $\lambda_{\rm eff}$,
rescaled by the bare Debye screening length (no added macro-ions), $ \lambda_{\rm D}$,
plotted as function of the macro-ion bulk concentration, $p_b$, rescaled by the bare salt concentration $n_0$.
The solid black curve is a symmetric case of $N_+=N_-=120$, the dotted red curve is a partially asymmetric case with $N_+=120$ and $N_-=60$,
and the dashed blue curve is a completely asymmetric case of  $N_+=120$  and $N_-=0$.
For all cases, $n_0 K_{\pm}=5$, where the chemical equilibrium constant, $K_\pm$, is defined in Eq.~(\ref{kinC}). Note that the peak in the symmetric case occurs at $p_b/n_0\simeq 1/N_\pm\simeq 8.3\times 10^{-3}$.}}
\label{Fig3}
\end{figure}

In Fig.~\ref{Fig3}, we plot $\lambda_{\rm eff}$ as a function of the macro-ion concentration, $p_b$ (at fixed salt concentration, $n_0$). The figure shows that when the macro-ion concentration is small, $\lambda_{\rm eff}$  increases as a function of $p_b$ for symmetric macro-ions, and decreases for asymmetric macro-ions, as explained above. However, at some point the increase/decrease of $\lambda_{\rm eff}$ reverses its sign, resulting in a non-monotonic dependence of $\lambda_{\rm eff}$ on $p_b$.

This changeover point occurs after a substantial portion of the cations/anions have been already adsorbed onto the macro-ions, such that the number of adsorbed cations/anions is comparable to the number of free cations/anions.
From this point on, each added macro-ion gets charged by removing cations or anions from the other macro-ions, and not only from the bulk, causing the number of cations and anions on each macro-ion to decrease. Asymmetric macro-ions become less charged, a process which increases $\lambda_{\rm eff}$ (dashed blue curve in Fig.~\ref{Fig3}). For symmetric macro-ions, the picture is more delicate: their overall charge remains zero, but the number of adsorbed cations and anions is reduced. Surprisingly, this results in a decreasing $\lambda_{\rm eff}$ (see Appendix~A for more details, and solid black curve in Fig.~\ref{Fig3}). Macro-ions that are partially asymmetric, meaning they adsorb both cations and anions but at different rates, display a combination of the symmetric and asymmetric behavior (dotted red curve in Fig.~\ref{Fig3}).
In general, unless $K_\pm$ have extremely small values, the crossover point occurs approximately at $p_bN_\pm \simeq n_0$ (where we assume that $N_+$ and $N_-$ are of the same order of magnitude).

\begin{figure*}
\subfloat{\includegraphics[width =
0.98\columnwidth,draft=false]{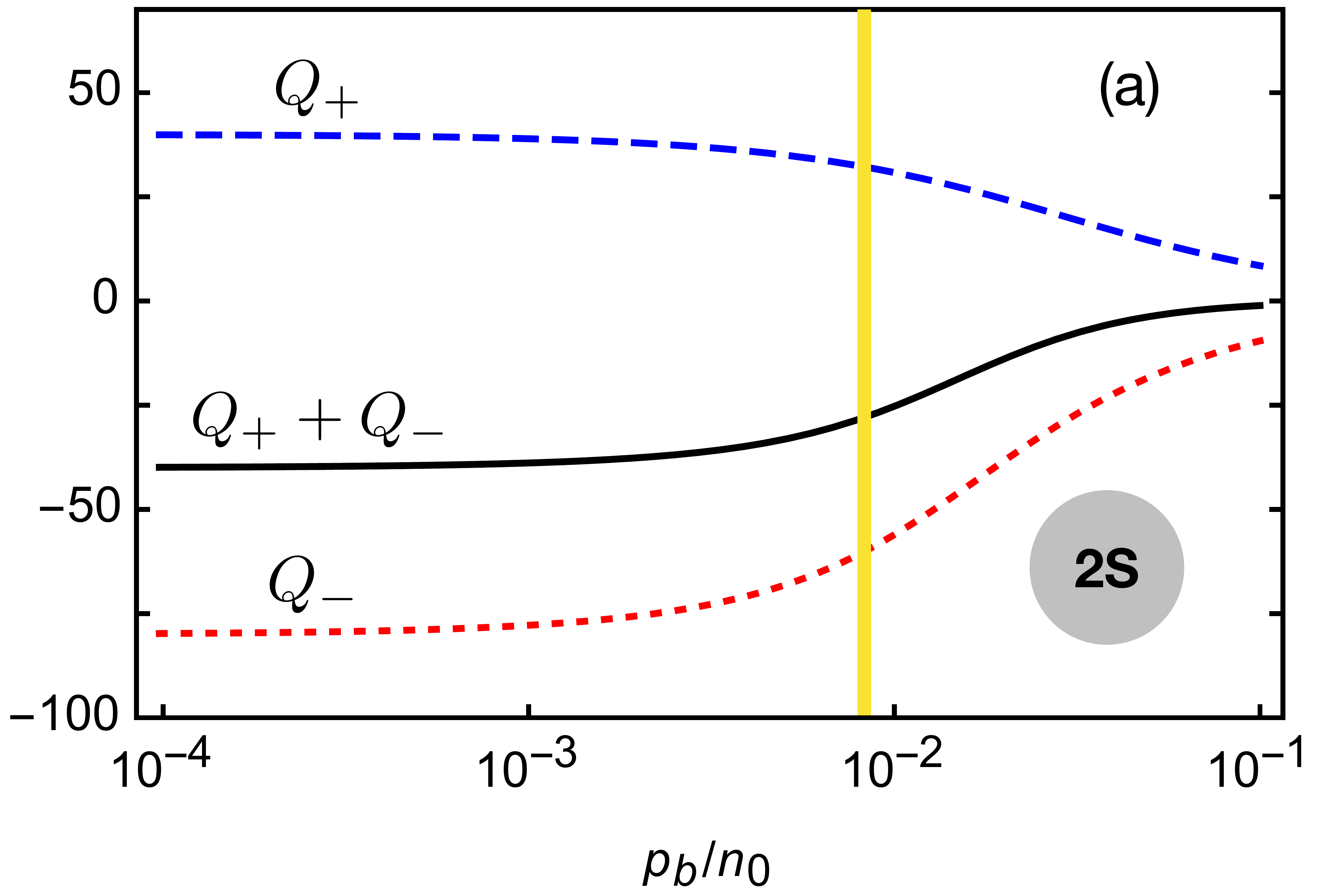}}\,\,\,\,\,\,\,\,\,\,\,\,\,\,\,
\subfloat{\includegraphics[width = 0.98\columnwidth,draft=false]{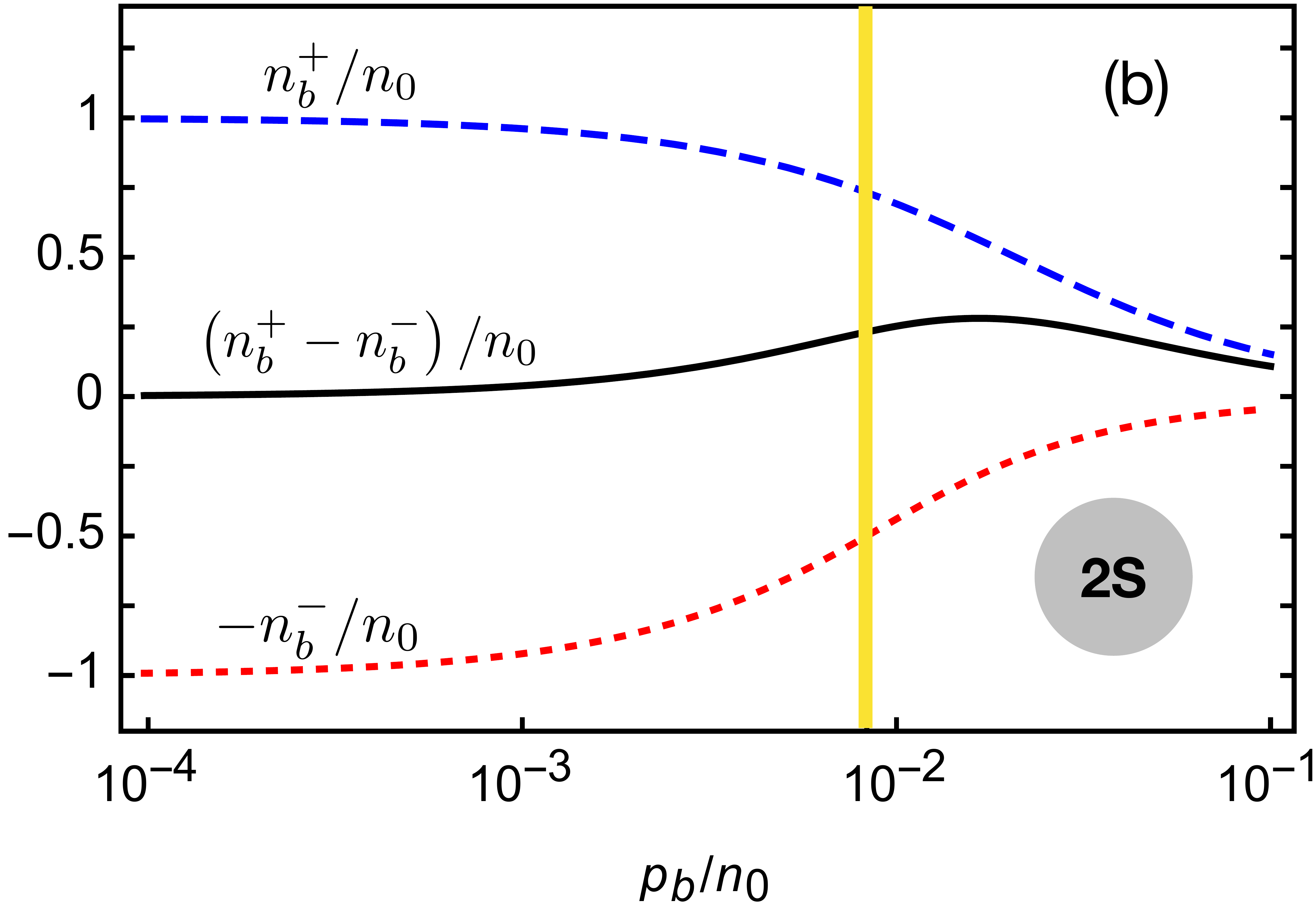}}
\caption{\textsf{(color online)  Results for the 2S model. (a) The macro-ion effective charge in the bulk as function of the macro-ion bulk concentration, $p_b$, rescaled by the bare salt concentration $n_0$. Parameter values are: $N_{\pm}=120$, $n_0K_{+}=1/2$, and $n_0K_{-}=2$. The dashed blue line corresponds to positive charge, $Q_{+}(0)=N_{+}\phi_{+}(0)$, dotted red line to negative charge, $Q_{-}(0)=-N_{-}\phi_{-}(0)<0$, and the solid black line to the overall effective charge, $Q(0)=Q_{+}(0) + Q_{-}(0)$. All charges are in units of $e$, and the vertical yellow line denotes the transition in behavior at $n_0=p_bN_\pm$. (b) The charge density of salt ions as function of the concentration of the macro-ions, $p_b$, rescaled by $n_0$, with the same parameters as in (a). All densities are rescaled by $n_0$. The dashed blue line corresponds to positive charge density, $n_b^{+}$, dotted red line to negative charge density, $-n_b^{-}$, and solid black line to overall charge density, $n_b^{+} - n_b^{-}$.}}
\label{Fig4}
\end{figure*}

The origin of the non-monotonic behavior is further manifested in Fig.~\ref{Fig4}. The effective macro-ion charge in the bulk and the salt ion bulk charge density are shown as a function of the rescaled macro-ion concentration, $p_b/n_0$. Figure~\ref{Fig4}(a) shows separately the positive charge, $Q_{+}(0) \equiv  eN_{+} \phi_{+}(0)$, the negative charge, $Q_{-}(0) \equiv - eN_{-} \phi_{-}(0)$, and the total charge, $Q(0)=Q_{+}(0)+Q_{-}(0)$, of the macro-ions in the bulk, Eq.~(\ref{ep2S}). Similarly, Fig.~\ref{Fig4}(b) shows separately (in units of $e$) the cation charge density $ n^{+}_b$, the anion charge density, $- n^{-}_b$, and the total charge density of the salt, $ n_b =  n^{+}_b - n^{-}_b$.
For small macro-ion concentrations, the positive and negative charge on each macro-ion barely change, and the change in the cation and anion densities is also  small. However, as soon as $p_{b}N_\pm \approx n_0$, there is a clear transition in behavior, characterized by the fact that the number of positive and negative charge per each macro-ion, as well as the cation/anion number density in the bulk, all sharply decrease, in accord with the above discussion.

\subsubsection{Symmetric macro-ions in presence of added salt}
\begin{figure}[h!]
\includegraphics[width=1\columnwidth,draft=false]{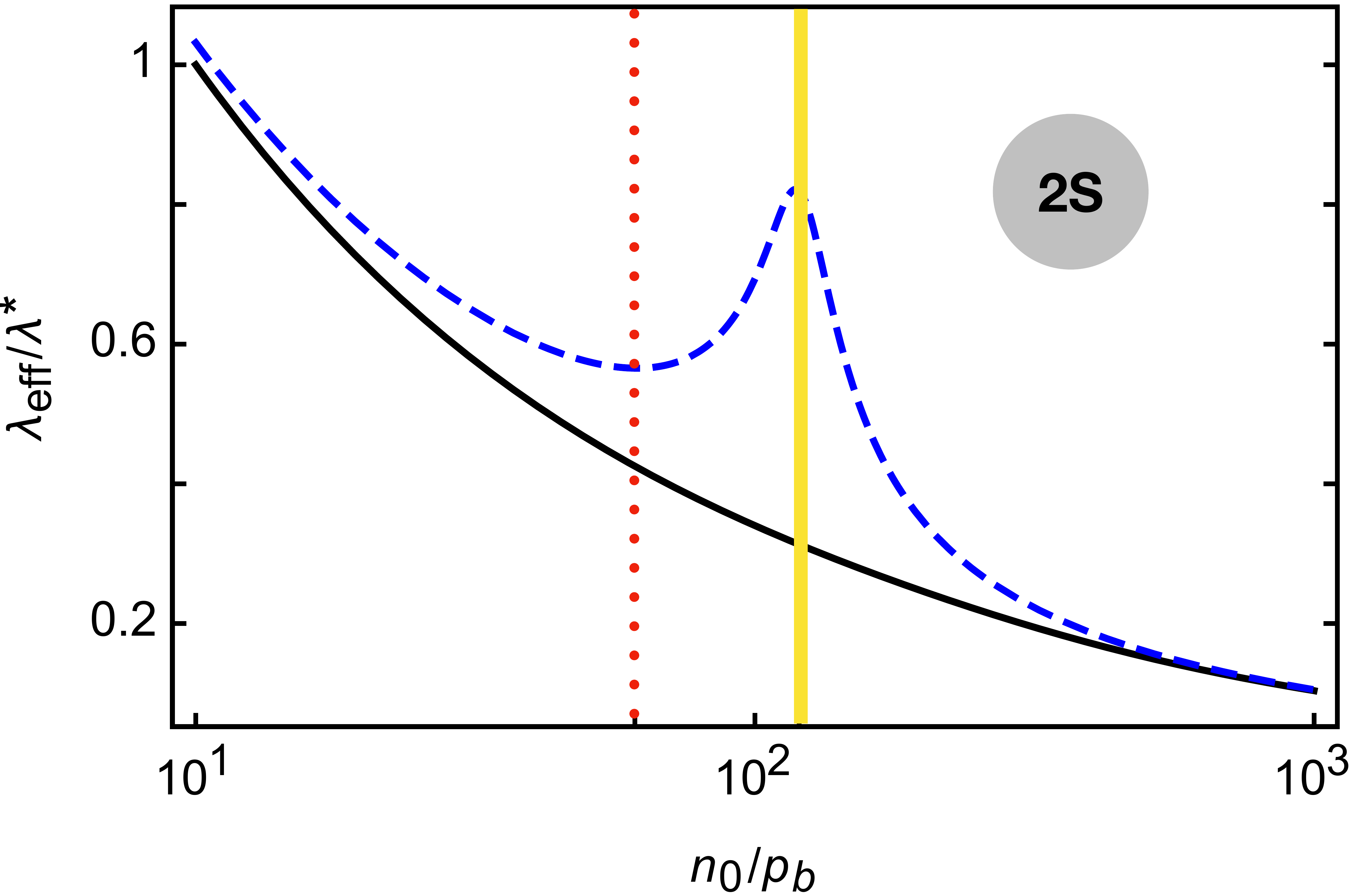}\centering
\caption{\textsf{(color online) $\lambda_{\rm eff}/\lambda^{*}$ for the 2S model, where $\lambda^{*}$  is the screening length of the solid black line at $n_0/p_b=10$, as a function of the concentration of extrenally added salt, $n_0$, rescaled by the fixed macro-ion concentration, $p_{b}$. The two curves represent symmetric macro-ions with $N_{\pm}=120$, which differ in their binding constants. Solid black line: $p_bK_\pm=1/100$, dashed blue line: $p_b K_\pm=2$.
The dashed blue curve has a minimum at $\phi_{\pm}(0)=1/2$ (vertical dotted red line) and a maximum at approximately $n_0=p_b N_{\pm}$ (vertical solid yellow line).}}
\label{Fig5}
\end{figure}

A rather non-intuitive phenomenon predicted by our 2S model can be seen in another way of changing system parameters. In Fig.~\ref{Fig5}, $\lambda_{\rm eff}$ is plotted in a symmetric case as a function of the concentration of externally added salt, $n_{\rm 0}$, at fixed $p_{b}$ in a symmetric case. The meaning of changing $n_0$ here and in Fig. 5 is that it allows us to control the overall salt concentration in the solution (adsorbed and bulk salt). When the chemical equilibrium constant is small ($p_bK_{\pm} \ll 1$, solid black line), $\lambda_{\rm eff}$  decreases monotonically and smoothly  as function of $n_0$. This decrease can be explained by the fact that for small $K_\pm$, the charging of macro-ions is small, and its effect on $\lambda_{\rm eff}$ is negligible compared to the decrease of $\lambda_{\rm eff}$ that comes from the additional salt.

However, when the chemical equilibrium constants are very large ($p_bK_\pm > 1$, dashed blue line), a different behavior is seen.
At small salt concentrations, the macro-ions adsorb an increasing number of cation/anion pairs. This results in a decrease of $\lambda_{\rm eff}$, followed by a sharp increase up to the saturation point in which $\lambda_{\rm eff}$ is maximal. This surprising phenomenon occurs because symmetric macro-ions screen the most when they are half-filled (see Appendix A for more details). The macro-ions decrease $\lambda_{\rm eff}$ until they are half-filled (vertical dotted red line in Fig.~\ref{Fig5}), and then increase $\lambda_{\rm eff}$ until their sites are completely occupied. The increase is dramatic because after all macro-ions are half-filled, the increase of salt concentration causes a {\em macroscopic} number of macro-ions to screen less  for a small addition of salt. The macro-ion charge saturates approximately at $p_b N_\pm = n_0$ (vertical solid yellow line in Fig.~\ref{Fig5}). From that point on, the added salt ions stay dissolved in the solution rather than adsorb on the macro-ions, causing $\lambda_{\rm eff}$ to drop. We see that due to the complexity of the macro-ion charging mechanism, addition of salt can in some cases increase $\lambda_{\rm eff}$.

\subsubsection{Donor vs. acceptor macro-ions}
Another case that deserves a separate discussion is the case where the macro-ions are added to the solution with already charged sites. This depends on the experimental initial conditions and how the addition of the macro-ions is done. We assume for simplicity that the number of initially occupied A sites equals the number of initially occupied C sites, and denote this number by $P_0$. Note that so far this number was assumed to be zero, $P_0=0$. The only change to the equations is a new electro-neutrality condition that now becomes
\begin{eqnarray}
\label{n_02S_with_P0}
n^\pm_{b} = n_{0}+p_{b}[P_0 - N_{\pm}\phi_\pm(0)].
\end{eqnarray}
It leads to an extra term in the $\lambda_{\rm eff}$ expression of Eq.~(\ref{screening2S})
\begin{eqnarray}
\label{screening2SP0}
\nonumber \lambda_{\rm eff}&=&  \lambda_{\rm D} \Bigg( 1 + \frac{p_{b}}{n_0}P_0 +
\frac{1}{2} \frac{p_{b}}{ n_0} \frac{Q^2(0)}{e^2} \\
&-& \frac{1}{2} \frac{p_{b}}{n_0} \Big[N_{+}\phi_{+}^{2}(0)+N_{-}\phi_{-}^{2}(0) \Big] \vphantom{\frac{A}{B}} \Bigg)^{-1/2} \, .
\end{eqnarray}

\begin{figure}[h!]
\includegraphics[width=1\columnwidth,draft=false]{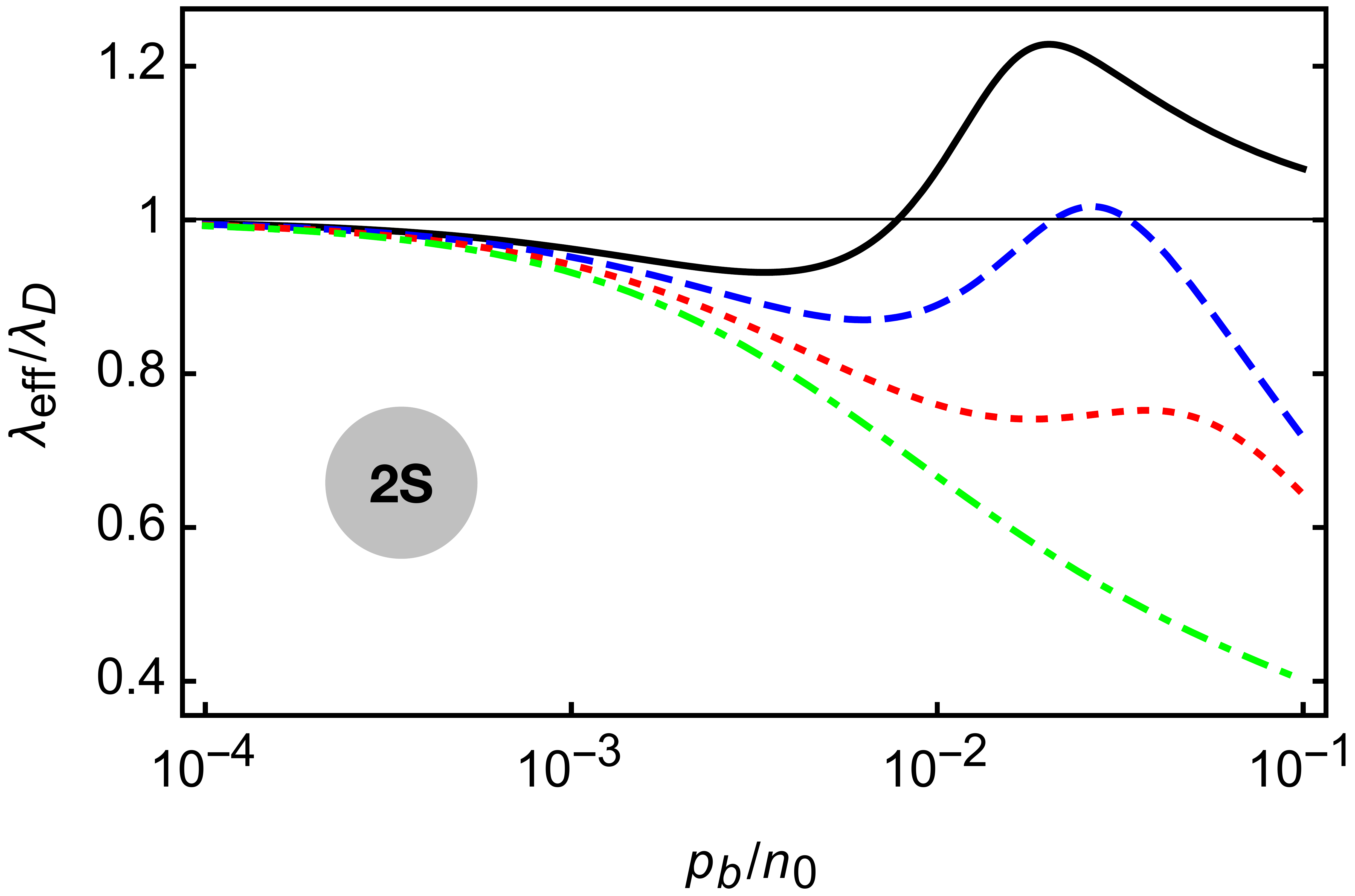}\centering
\caption{\textsf{(color online) The screening length in the 2S model, $\lambda_{\rm eff}/\lambda_D$, as function of the macro-ion bulk concentration, $p_b/n_0$, for different $P_0$ values. Parameter values are: $N_+=100$, $N_-=60$, $n_0K_+=2$ and $n_0K_-=5$. From top to bottom to bright: $P_0=0$ (solid black line), $P_0=20$ (dashed blue line), $P_0=40$ (dotted red line) and $P_0=60$ (dot-dashed green line). The increase of $P_0$ decreases the screening length because effectively more salt is added to the solution.}}
\label{Fig6}
\end{figure}

In Fig.~\ref{Fig6}, we plot $\lambda_{\rm eff}$ as a function of the macro-ion concentration, $p_b$, for different values of $P_0$. As $P_0$ is increased, the screening length decreases. In addition, the non-monotonic behavior of $\lambda_{\rm eff}$ diminishes and eventually disappears. This is due to the fact that at large $P_0$ values, the addition of macro-ions comes with the addition of large salt concentration which overcomes the unique macro-ion effects, and decreases $\lambda_{\rm eff}$ in an obvious way.

Note that when $P_0$ is roughly larger than both $N_{\pm}\phi_{\pm}$, each macro-ion releases more ions than it adsorbs. In this case the macro-ions are classified as {\it donors}. For the opposite case of $P_0<N_{\pm}\phi_{\pm}$, the situation is reversed, and the macro-ions can be classified as  {\em acceptors}. Donor macro-ions are more likely to decrease the screening length than acceptors, while the screening length shows non-monotonic dependence on the macro-ion concentration only for acceptors. The donor/acceptor classification depends on the way the macro-ions were introduced initially into the solution, and is characterized by their $P_0$ charge parameter.

\subsection{The one-site (1S) model}

Next, we consider a CR model that is referred to as the 1S model because the macro-ions have only one type of active sites. These sites are first taken as electroneutral, but they can either desorb or adsorb  a cation, with a different free-energy cost. In units of the elementary charge $e$,  the overall charge of the site gets one of the three values: $-1$, $0$ or $+1$ (see Fig.~\ref{Fig2}(b)).
This specific adsorption process  involves just a single type of solution ion (positive). It can be described by two chemical reaction equations:
\begin{equation}
\label{reaction1S}
\begin{split}
\text{AB} & \rightleftarrows \text{A}^{-} + \text{B}^{+}\\
\text{AB}_2^{+} & \rightleftarrows \text{AB} + \text{B}^{+} \,.
\end{split}
\end{equation}
Such dissociation equilibrium can be used to model amphoteric (zwitterionic) charge processes~\cite{ettelaie1995electrical, chan1976electrical, biesheuvel2004electrostatic}. It  was studied experimentally
in Ref.~\cite{morag2013governing}, on silica surfaces containing SiO$^{-}$ groups that adsorb either one $\text{H}^{+}$ ion or two $\text{H}^{+}$ from the solution, and become $\text{SiOH}$ or $\text{SiOH}_2^{+}$, respectively.

The total number of active sites is $N$, and we define the fraction of positively (negatively) charged sites $\phi_{+}$ ($\phi_{-}$). According to our model, $\phi_{-}$ is the fraction of sites that have released one cation to the solution, $1-\phi_{+} - \phi_{-}$ is the fraction of neutral sites, and $\phi_{+}$ is the fraction of sites that adsorbed one cation from the solution. Therefore, the overall number of potentially dissociable positive ions on a macro-ion is  $(1 -  \phi_{+} - \phi_{-}) + 2\phi_{+}=1+\phi_{+} - \phi_{-}$. The free-energy gain of a neutral site adsorbing a cation is $\alpha_{+}$, while the free-energy gain of a negatively charged site adsorbing a cation is $\alpha_{-}$.

The free energy of a single macro-ion is then given by,
\begin{equation}
\label{g1S}
\begin{split}
g^{\rm 1S}\left(\psi,\phi_{+},\phi_{-}\right) & = N\bigl(\phi_{+}-\phi_{-}\bigr)e\psi \\
&- N\bigl(\phi_{+}\alpha_{+}-\phi_{-}\alpha_{-}\bigr) \\
& -\, N\left(1+\phi_{+}-\phi_{-}\right)\mu_{+}\\
 & +\, k_{B}TN\Bigl[\phi_{+}\ln\phi_{+} + \phi_{-}\ln\phi_{-} \\
 & +\, \left(1-\phi_{+}-\phi_{-}\right)\ln\left(1-\phi_{+}-\phi_{-}\right)\Bigr] .
\end{split}
\end{equation}
Two differences can be observed between the free energies of the 2S and 1S models, $g^{\rm 2S}$ and $g^{\rm 1S}$. The first one is the Lagrange multiplier $N\left(1+\phi_{+}-\phi_{-}\right)\mu_{+}$ term in Eq.~(\ref{g1S}), which corresponds to the fact that cations are exchanged with the solution.
The second difference is in the entropic part. It is modified because there is only one type of active sites, and each site in the 1S model can be in three states: positive, negative or neutral.

The variation of $F$, Eq.~(\ref{bp1}), is now straightforward and is done using Eq.~(\ref{g1S}), along the same lines as for the 2S model above. We obtain,
\begin{equation}
\label{phi1S}
\phi_{\pm}(\psi) =\frac{Z^{\rm{1S}}_{\pm}(\psi)-1}{Z^{\rm{1S}}_{+}(\psi)+Z^{\rm{1S}}_{-}(\psi)-1},
\end{equation}
where
\begin{eqnarray}
\label{Z1S12}
Z^{\rm 1S}_{+}&=&1+(n_{b}^{+} K_{+})\,{\rm e}^{- \beta e\psi}\nonumber\\
Z^{\rm 1S}_{-}&=&1+(n_{b}^{+}K_{-})^{-1}\,{\rm e}^{\beta e \psi},
\end{eqnarray}
depend on the appropriate chemical equilibrium constants, $K_\pm=a^3\exp(\beta\alpha_\pm)$, for the reactions as in Eq.~(\ref{reaction1S}).
$K_{+}$ is the chemical equilibrium constant associated with  a neutral AB site binding a cation (B$^+$) and becoming AB$_2^{+}$,
while $K_{-}$ is associated with A$^{-}$ adsorbing a cation (B$^+$) and becoming a neutral AB.
For simplicity, we omit throughout this subsection the superscript 1S.

Substituting the expression for $\phi_\pm$ [Eq.~(\ref{phi1S})] in Eq.~(\ref{g1S}), we get $g(\psi)$ in the form
\begin{equation}
\label{g1S2}
g(\psi) =-Nk_{B}T\ln\left[Z_{+}(\psi)+Z_{-}(\psi)-1\right],
\end{equation}
where we omitted a constant term that has no thermodynamic consequences. Substituting Eq.~(\ref{g1S2}) into Eq.~(\ref{p}), gives the macro-ion concentration as
\begin{equation}
\label{p1S}
p(\psi) = p_{b}\left(\frac{Z_{+}(\psi)+Z_{-}(\psi)-1}{Z_{+}(0)+Z_{-}(0)-1}\right)^{N}
\, .
\end{equation}
The charge of each macro-ion for the 1S model is then
\begin{eqnarray}
\label{epS1}
\nonumber Q(\psi) &=& eN \left(\phi_{+} - \phi_{-}\right) \\
&=& eN\left(\frac{Z_{+}(\psi)-Z_{-}(\psi)}{Z_{+}(\psi)+Z_{-}(\psi)-1}\right) \, ,
\end{eqnarray}
while the total charge density of macro-ions and salt monovalent ions is obtained by inserting Eqs.~(\ref{p1S}) and (\ref{epS1}) into Eq.~(\ref{PB}).

The relation between $n_{b}^\pm$ and $n_0$  and $p_b$  ensures an overall electro-neutrality
\begin{eqnarray}
\label{neutralityS1}
\label{n_01S}
n^{+}_{b} &=& n_{0}-p_{b}N\left(\phi_{+}-\phi_{-}\right)\nonumber \, ,\\
n^{-}_{b} &=& n_{0} \, .
\end{eqnarray}
The above equality, $n^{-}_{b} = n_{0}$, is valid because in the 1S model we explicitly assumed that the anions do not participate in the CR process of the macro-ions.\footnote{In contrast to the 2S model, even if some sites are initially charged, there is no need to introduce $P_0$. As long as the macro-ion is initially overall neutral, it has $N $ potentially dissociable ions that can be released into the solution, regardless of its initial occupancy, and Eq.~(\ref{neutralityS1}) holds.}

\subsection{Screening length for 1S model} \label{Screening section1}

We now discuss the screening phenomenology in the 1S model. Substituting
Eqs.~(\ref{rho2S}) and~(\ref{n_01S}) into Eq.~(\ref{screening}) results in the screening length of the form
\begin{equation}
\lambda_{\rm eff}=\frac{\lambda_{\rm D}}{\left(1+\frac{1}{2}\frac{p_{b}N}{n_0}\left[\left(N-1\right)
\left(\frac{Q(0)}{e}\right)^2 + 2\phi_{-}(0)\right]\right)^{1/2}}\, ,
\label{screening1S}
\end{equation}
recalling that $Q(0)$ and $\phi(0)$ are bulk values taken at zero potential, $\psi=0$.
One immediate consequence is that  $\lambda_{\rm eff}$ can only decrease relative to $\lambda_{\rm D}$, to be distinguished from the 2S model, where both increase and decrease of $\lambda_{\rm eff}$ are possible. More specifically, in the 1S model the macro-ion {\em cannot} adsorb pairs of positive and negative ions, which is the mechanism allowing an increase in the screening length in the 2S model.

The term proportional to $Q(0)$ represents the charge asymmetry, and the larger it is, the smaller is the screening length. However, to obtain full symmetry, {\em i.e}, $Q(0)=0$, the relation between the different constants is not as simple as in the 2S model. According to Eq.~(\ref{epS1}),  the condition is that $Z_{+}$  and $Z_{-}$, defined in Eq.~(\ref{Z1S12}) , should be equal to one another at $\psi=0$. It then follows that there is no choice of $K_{+}$ and $K_{-}$ which guarantees $Q(0)=0$,
regardless of the salt bulk concentration. Hence, the symmetry between positive and negative sites for a given bulk concentration does not guarantee that such a symmetry will hold also for a different concentration.

As for comparing donor and acceptor features, one should keep in mind that in the 1S model each macro-ion has initially (before it is added to the solution) $N$ monovalent bound cations that can be released.  Once placed in the solution, it can adsorb or release cations and remain with any number of them between 0 to $2N$. From the definition of $\phi_{+}$ and $\phi_{-}$ in the 1S model, it follows that if $\phi_{-}>\phi_{+}$, then the macro-ion is a donor, while if $\phi_{-}<\phi_{+}$, it is an acceptor. The last term in the denominator of Eq.~(\ref{screening1S}), proportional to $\phi_{-}(0)$, is related to this feature: the larger $\phi_{-}(0)$ is, the more likely it is that the macro-ion behaves as a donor and decreases $\lambda_{\rm eff}$.

\begin{figure}
\includegraphics[width=1\columnwidth,draft=false]{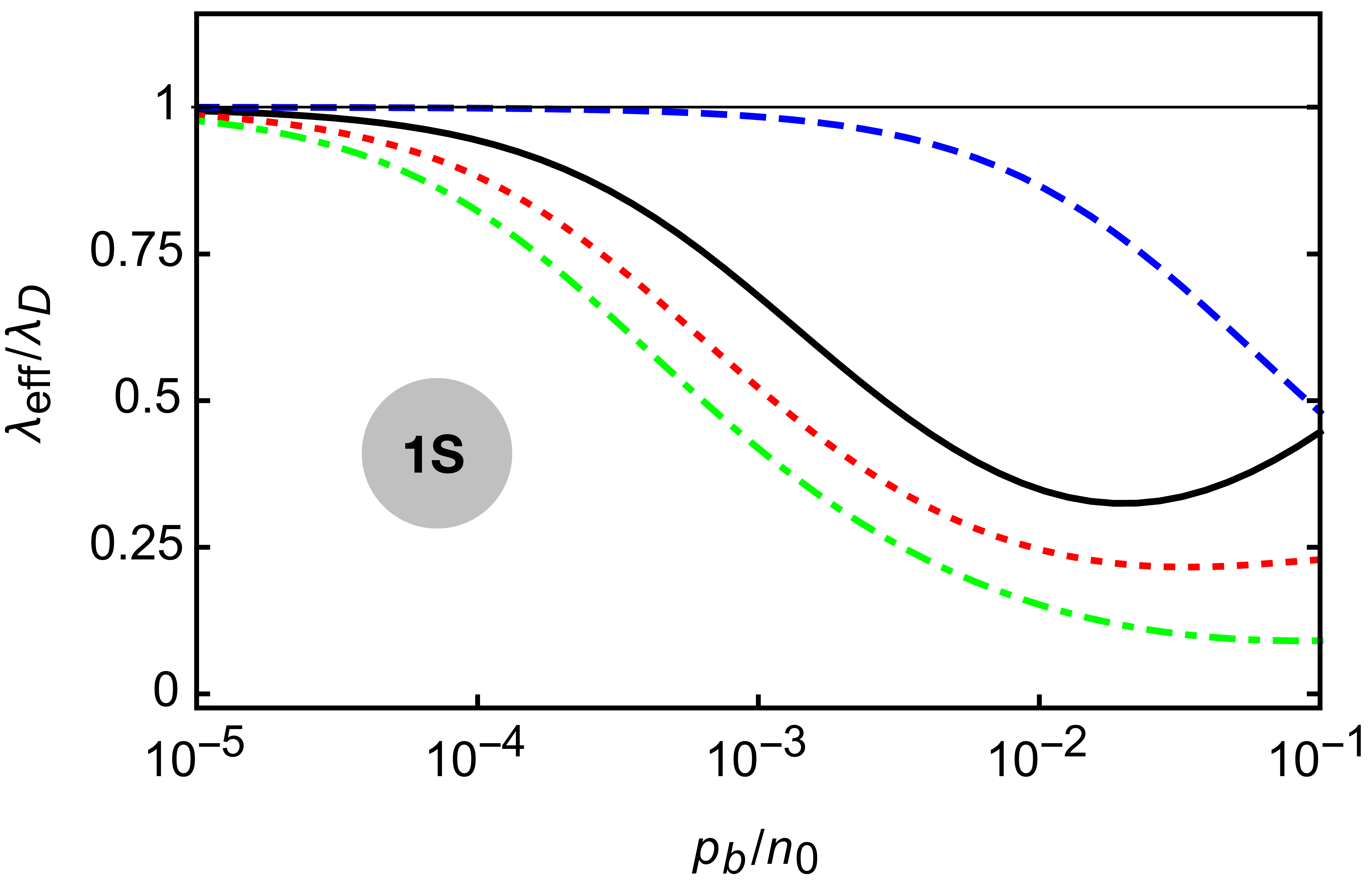}\centering
\caption{\textsf{(color online) Results for the 1S model. The screening length, $\lambda_{\rm eff}$, rescaled by $ \lambda_{\rm D}$, the bare Debye screening length (no added macro-ions), plotted as function of the macro-ion bulk concentration, $p_b$, rescaled by the bare salt concentration $n_0$.
 The four curves from top to bottom are: $n_0 K_{-}=1$ (dashed blue line), $K_{-}\to \infty$ (solid black line), $n_0 K_{-}=1/10$
 (dotted red line), and $n_0 K_{-}=1/100$ (dot-dashed green line).
 For all curves,  $N=100$ and $n_0 K_{+}=1$.}}
\label{Fig7}
\end{figure}

In Fig.~\ref{Fig7},  $\lambda_{\rm eff}$ for the 1S model is plotted as a function of $p_b$ (fixed $n_0$).
The plotted blue, red and green curves show a regular behavior. The addition of macro-ions decreases the screening length, similar to addition of regular ions. This is attributed to the fact that in these cases the macro-ions donate a large amount of ions to the solution. The solid black line describes a different case where neutral macro-ion sites cannot dissociate ($K_{-} \to \infty$). Therefore, they behave as acceptors, and display a non-monotonic behavior similar to acceptor macro-ions in the 2S model.
\begin{figure*}
\subfloat{\includegraphics[width = 1\columnwidth,draft=false]{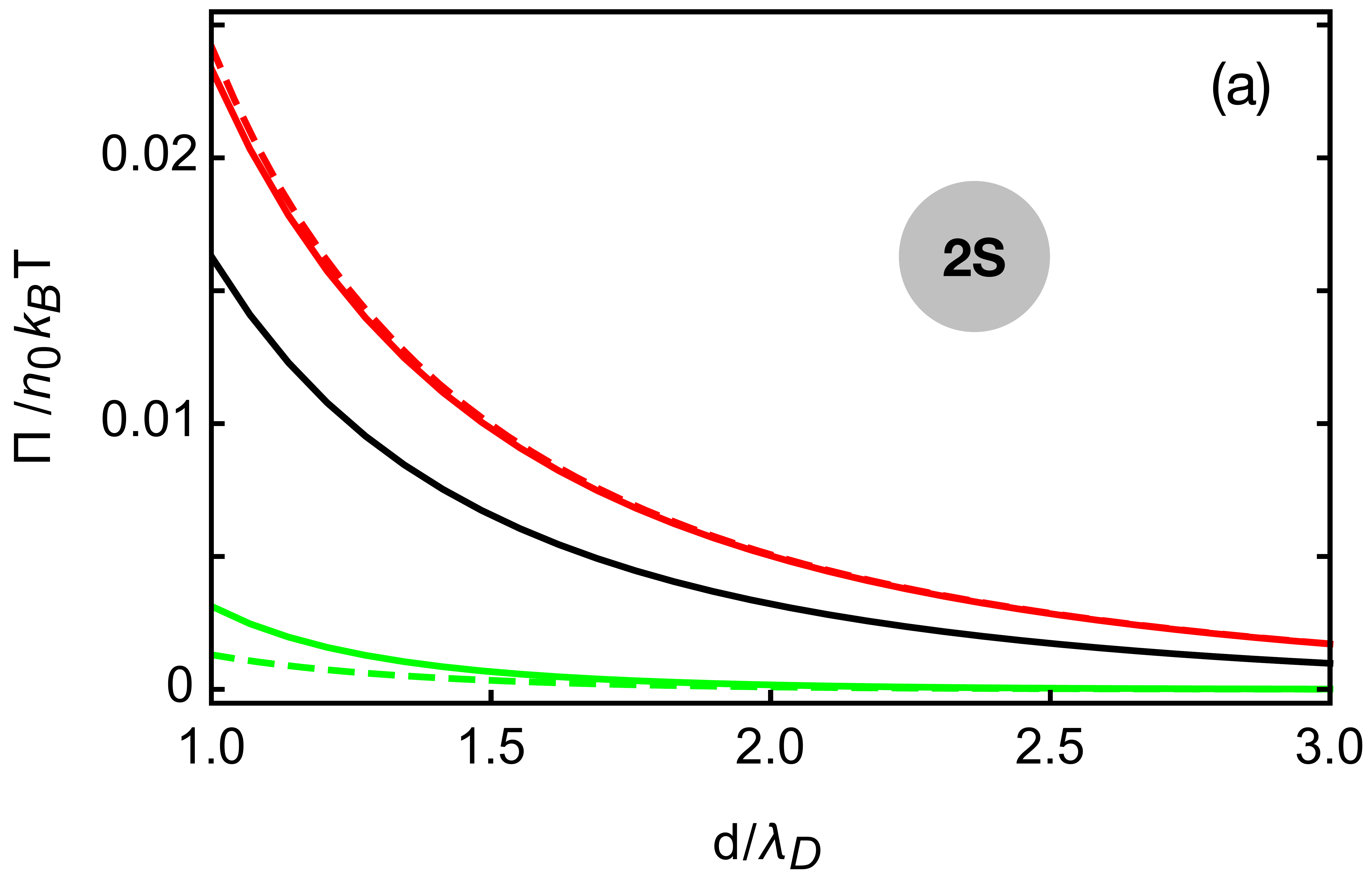}}
\subfloat{\includegraphics[width = 1.\columnwidth,draft=false]{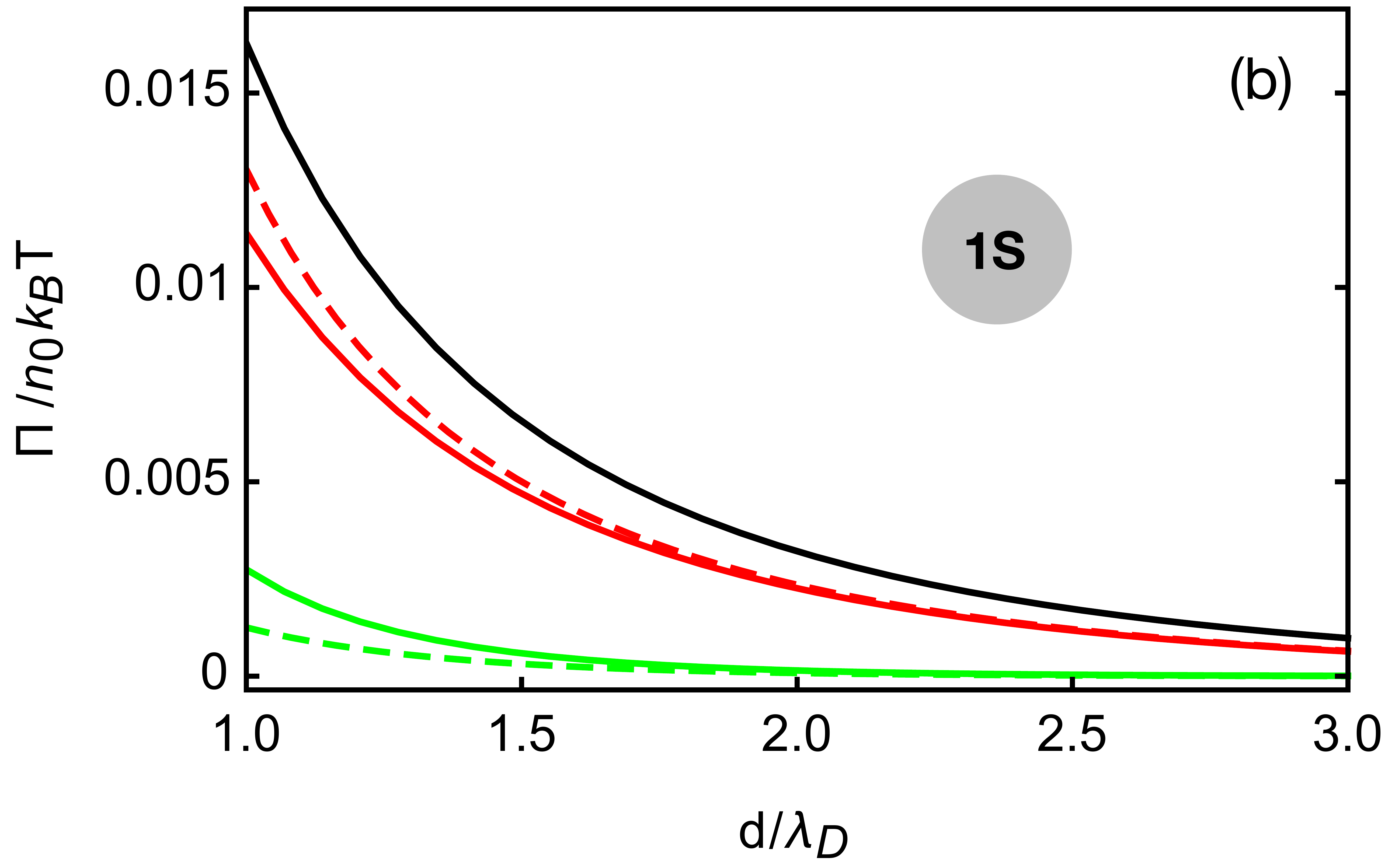}}
\caption{\textsf{(color online)  Disjoining pressure $\Pi$ (rescaled by $n_0 k_{B}T$), as function of the distance between the surfaces, $d$ (rescaled by $\lambda_{\rm D}$). Solid lines correspond to exact PB theory, while dashed lines to the DH approximation, where the electrostatic potential is given by
$\psi(z)=-(2\lambda_{\rm eff}/l_{\rm GC}) \cosh (z/\lambda_{\rm eff})/\sinh (d/2\lambda_{\rm eff})$.
(a) Results for the 2S model, for the asymmetric case, $K_{+}\to 0$ and $n_0 K_{-}=10$ (green lines),
and the symmetric case, $n_0 K_\pm=10$ (red lines). The black corresponds to no added macro-ions,
$p_b=0$, and is plotted for comparison. The other parameter values are: $l_{\rm GC}/\lambda_{\rm D}=30$, $p_b/n_0=0.01$, $N_\pm=40$ and $P_0=0$.
(b) The same plot as in (a), but for  the 1S model. Parameter values are: $N=80$, $K_{+} \to \infty$ and $n_0K_{-}=1$
(green lines), and $n_0 K_\pm=1$ (red lines). The parameters $l_{\rm GC}/\lambda_{\rm D}$ and $p_{b}/n_0$ have the same values as in (a).}}
\label{Fig8}
\end{figure*}

\subsection{Disjoining pressure between two charged surfaces}
We  examine how the force (or equivalently the
disjoining pressure) between two charged planar surfaces, bearing fixed surface charge density,  is modified when they are embedded in a salt solution containing in addition charge-regulating macro-ions.

The disjoining pressure between two symmetric surfaces placed at $z{=}\pm d/2$,  and each carrying a fixed charge density $\sigma$, is calculated by integrating the PB equation, Eq.~(\ref{PB}), with the boundary conditions, ${\rm d}\psi/{\rm d}z |_{z=\pm d/2}=\pm  \sigma /(\varepsilon_0 \varepsilon)$.
In fact, the pressure between the surfaces is known to be~\cite{Safinya,Maggs2016general},
\begin{eqnarray}
\label{OsmPress}
\Pi  =&-&\frac{\varepsilon_0 \varepsilon}{2}\psi'^{2}(z)
+ k_{B}T \sum_{i=\pm}\Bigl[n_{i}(z) - n^{i}_{b}\Bigr] \nonumber \\
&+&\,k_{B}T\Bigl[p(z)-p_{b}\Bigr]\, .
\end{eqnarray}
This pressure is a constant in equilibrium, independent of $z$, although each of its three terms is $z$-dependent. For convenience, we choose $z$ to be at the midplane, $z=0$, where by symmetry the electric field, $\psi'(0)=0$, in Eq.~(\ref{OsmPress}), yielding a simplified form for $\Pi$,
\begin{equation}
\Pi= k_{B}T \sum_{i=\pm} \,( n_{i}(0) - n^{i}_{b} )+k_{B}T(p(0)-p_b) \, .
\end{equation}
The disjoining pressure $\Pi$ is then reduced to the difference in the ideal (van 't Hoff) osmotic pressure of three ionic species, calculated between the point at the midplane and the bulk.

In the Debye-H\"uckel (DH) limit, valid for $\lambda_{\rm eff}^{2}/l_{\rm GC}d \ll1$, with the Gouy-Chapman length defined as $l_{\rm GC}=2\varepsilon_0 \varepsilon k_{B} T/ e \left|\sigma\right|$, the electrostatic potential $\psi$ between two charged surfaces is obtained from the linearized PB equation, Eq.~(\ref{PB1}), and depends only on the effective screening length, $\lambda_{\rm eff}$~\cite{Safinya}.
The electrostatic potential determines the salt ion and macro-ion concentrations, which in turn determine $\Pi$. Consequently, we expect the disjoining pressure to exhibit, a non-monotonic dependence on system parameters, stemming from the $\lambda_{\rm{eff}}$ behavior. Intuitively, we can argue that the screening decreases the range of the interaction between the two surfaces, and in turn this decreases the disjoining pressure.

In Fig.~\ref{Fig8}, we plot the disjoining pressure, $\Pi(d)$, between the two charged surfaces, as function of $d$, the inter-plate distance. The pressure dependence is obtained by solving numerically the full PB equation for the 2S and 1S models, and, for comparison we also plot $\Pi(d)$ for the DH approximation (dashed lines). As is clear from the figure, the disjoining pressure always decreases monotonically with $d$, just as is expected for simple ionic solutions~\cite{Safinya}. We note that for large $d$, where the electrostatic potential is small, the full PB and the DH curves coincide.
For the 2S model, the addition of macro-ions can either increase or decrease $\Pi$ for all values of $d$,
as can be seen by comparing the plots with macro-ions (red and green) with the curve without them (black).
This agrees with the result obtained in Sec.~\ref{Screening section}, since $\lambda_{\rm eff}$ can increase or decrease, as function of the concentration of the macro-ions, $p_b$. On the other hand, in the 1S model, $\lambda_{\rm eff}$ always decreases (see Sec.~\ref{Screening section1} and Fig.~\ref{Fig7}), so the resulting $\Pi$ always decreases as function of $p_b$, for all values of $d$. Although not plotted, for each distance $d$ between the surfaces, the density profiles $n_{\pm}(z)$, $p(z)$ and the charging profiles $Q(z)$, $\phi_{\pm}(z)$ can be easily obtained from our calculations.

\begin{figure}[h!]
\includegraphics[width=1\columnwidth,draft=false]{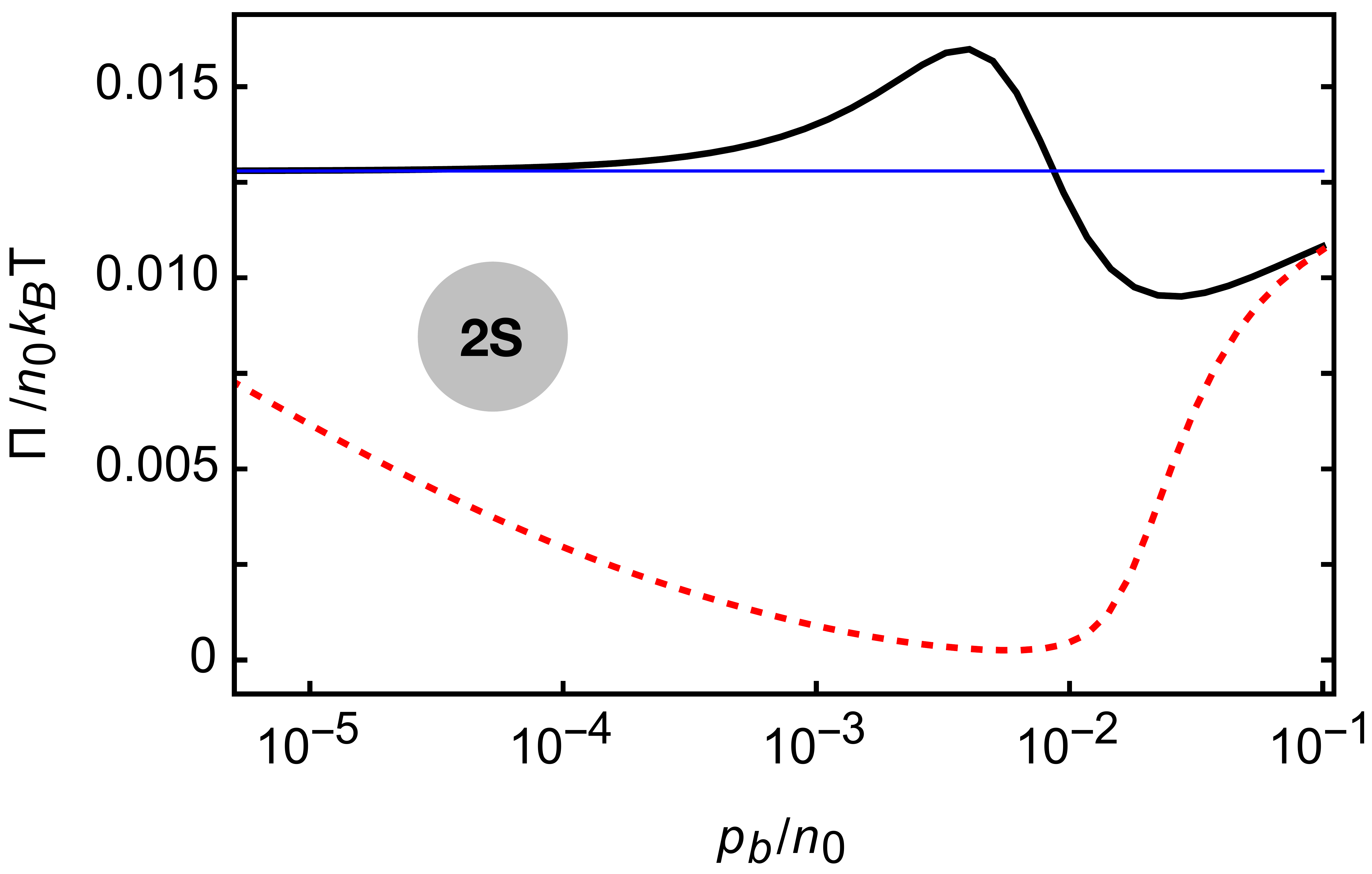}\centering
\caption{\textsf{(color online)  Disjoining pressure $\Pi$ in the 2S model (rescaled by $n_0 k_{B} T$), as function of the macro-ion concentration $p_{b}$ (rescaled by the bare salt $n_0$). Parameter values are:  $N_{+}{=}120$, $N_{-}{=}120$ (solid black line), and $N_{+}{=}120$, $N_{-}{=}60$ (dotted red line). The reference pressure of no macro-ions is shown as the thin blue line. Other parameter values are: $d/\lambda_{\rm D}=2$,  $d/l_{\rm GC}=2/15$,   $n_0 K_\pm=5$ and $P_0=0$.}}
\label{Fig9}
\end{figure}

We conclude this section by showing in Fig.~\ref{Fig9} the disjoining pressure in the 2S model for a fixed value of the inter-plate distance $d$ as function of $p_b/n_0$, for different charge asymmetry cases (as defined in Sec.~\ref{Screening section}). The macro-ion parameters in Fig.~\ref{Fig9} are the same as for the solid black and dotted red lines in Fig.~\ref{Fig3}.
At low macro-ion concentration, symmetric macro-ions increase the inter-plate pressure, in agreement with the results presented in Fig.~\ref{Fig8}, while more asymmetric macro-ions decrease it. When $N_\pm p_b \approx n_0$,  a sharp change in the behavior is seen and the original trend changes its  sign. This agrees with the $\lambda_{\rm eff}$ behavior depicted in Fig.~\ref{Fig3}. In the symmetric case, another sharp change is seen for higher macro-ion concentration, where the pressure begins to increase again. This changeover cannot be simply understood by the screening length behavior, but involves non-linear effects due to the charged boundaries. Therefore, the charge regulation mechanism contains further rich phenomena beyond what is  explored in this paper, especially those related to higher-order effects.

\section{Conclusions} \label{discussion}

In this work we have addressed the role of mobile and complex charge-regulating (CR) macro-ions dissolved in a simple salt solution.
Our goal  is to highlight the collective effects, and to treat on par the translational degrees of freedom of the CR macro-ions with those of the simple salt ions~\cite{Markovich2017EPL}. This more general treatment leads to quite unexpected outcomes, and points out to  rich patterns exhibited by  complex colloid solutions.
In particular, we scrutinized the resulting screening properties as well as the interactions between two charged interfaces immersed in CR colloid solution.

Two CR models are studied in detail: the two-site (2S) and one-site (1S) models.
The 2S model has two uncorrelated sites.
One of them can adsorb a cation and the other can adsorb an anion. The situation is somewhat different in the 1S model, where there is only one type of sites that can either adsorb or desorb a cation.
As can be seen in this work, the two models give quite different results.

Our most important finding is the anomalous non-monotonic dependence of the effective screening length ($\lambda_{\rm eff}$) on the bulk concentrations of the macro-ions $p_b$, and the salt $n_0$. As can be seen in Figs.~\ref{Fig3} and~\ref{Fig5}-\ref{Fig7}, the screening displays even more unexpected features, in addition to those already identified for a variant of the 2S model in Ref.~\cite{Markovich2017EPL}. Besides, the non-monotonic screening is reflected also in the behavior of the disjoining pressure, $\Pi$, between two planar surfaces with fixed charges, as function of the macro-ion concentration, $p_b$, at a fixed value of the inter-surface separation, $d$, as is clear from Fig.~\ref{Fig9}.

A detailed analysis shows that for the 2S and 1S models the dependence of $\lambda_{\rm eff}$ on the salt and/or macro-ion concentrations is governed by two important characteristics of the macro-ion CR process: (i) the balance between the different association processes, which can be quantified by whether the macro-ion prefers to have a neutral  overall charge (symmetric state),
or a non-zero value (asymmetric state), and (ii) the donor/acceptor propensity that relates to the macro-ion preference to either release ions  to the bathing solution, or acquire ions from it (and sometimes, it also depends on initial solution preparation conditions).

In both 1S and 2S models, macro-ions that prefer an overall non-zero charge ($Q\ne 0$), {\em i.e.,} exhibiting an {\em asymmetric} charging process, display increased screening (smaller $\lambda_{\rm eff}$). Consequently, they reduce the disjoining pressure between the  charged bounding surfaces when compared with a system of no added macro-ions at the same solution conditions.
Contrary, {\it symmetric} macro-ions prefer an overall close to zero charge ($Q \approx 0$). In the 2S model, this is achieved by adsorption of cation-anion pairs, which increases $\lambda_{\rm eff}$. In the 1S model, it is achieved by releasing and adsorbing the same amount of positive ions, still decreasing $\lambda_{\rm eff}$, but to a lesser extent than for the asymmetric case.

Acceptor macro-ions, common to the 1S and 2S models tend to adsorb more cations and anions from the bathing solution than they release. They present a more interesting case than the opposite kind of macro-ions (donors). Increasing the acceptor concentration leads to a transition from a state where most of the ions are in the bulk, to a state where most of them are adsorbed by the macro-ions. This transition leads to non-monotonic dependence of the screening length, $\lambda_{\rm eff}$, on the macro-ion bulk concentration. In addition, donor/acceptor propensity can either relate to the experimental initial condition ($P_0$ parameter in the 2S model), or it can be an inherent macro-ion property (like in the 1S model).

In the 2S model, a further unexpected phenomenon is observed in the screening properties of symmetric macro-ions.  At first glance, one may assume that such macro-ions do not contribute to the screening, as they carry an overall zero charge at zero electrostatic potential. However, we demonstrate that this is not the case, and further show that the macro-ions have small contribution to the screening for zero or full occupation, while they have a large contribution for half occupancy (see Eq.~(\ref{variance}) and Fig.~\ref{Fig5}).
This property has a novel non-intuitive implication. In extreme cases, it causes the screening length to exhibits a rapid increase with the addition of salt.
This behavior, besides being contrary to the behavior expected for the regular $\lambda_{\rm D}$, exhibits some features similar to a phase transition, in the sense that a small change in the system parameter ($n_0$) inflicts a macroscopic response of $\lambda_{\rm eff}$.

\bigskip\bigskip
Quite recently,  some experiments have been conducted on charge regulation of non-polar colloids~\cite{Bartlett}, and are pertinent to our analysis~\cite{Markovich2017EPL}. In the experimental system, the colloid macro-ions are dispersed in hydrophobic dodecane (non-polar) solvent and carry a charge that can self-adjust due to dissociation/association of  weakly-ionized surface groups. The effective charge of the colloid macro-ions is estimated from fits to scattering experiments of concentrated colloidal dispersions, and one of the main findings is a non-monotonic behavior as a function of the colloidal concentration. While in a broad sense, this experimental system is close to our theoretical ramifications, the details of the charging mechanism and the relation between the experimentally determined effective charge and our own effective colloidal charge, $Q$, are at present not entirely sorted out, and are left for follow-up studies.

In this work, we advocate a line of reasoning that charge regulation is essential in modeling complex macro-ion dispersed in  ionic solutions.
More detailed quantitative comparison between theory and experiment requires further understanding of the charging mechanism in experiments. These as well as specific protein segregation and adsorption phenomena, will be investigated in future studies.

\bigskip\bigskip
{\em Acknowledgements.}~~
We would like to thank R. Adar, M. Biesheuvel, D. Harries and Y. Tsori for useful discussions and numerous suggestions. This work was supported in part by the ISF-NSFC (Israel-China) joint research program under Grant No. 885/15. TM acknowledges the support from the Blavatnik postdoctoral fellowship programme, and RP was supported by the 1000-Talents Program of the Chinese Foreign Experts Bureau. RP would like to thank the School of Physics and Astronomy at Tel Aviv University for its hospitality, and is grateful for the Sackler Scholar fellowship and the Nirit and Michael Shaoul fellowship, awarded
within the framework of the Mortimer and Raymond Sackler Institute of Advanced Studies at Tel Aviv University.

\appendix
\section{Electrostatic screening by symmetric macro-ions (2S model)}
We concentrate on symmetric (neutral on average) macro-ions and show how in the 2S model  their screening length depends on the number of cation/anion pairs that the macro-ions adsorb.
We assume complete symmetry: $K_{\pm}=K$, $N_{\pm}=N$, $n_{b}^{\pm}=n_{b}$, and fix the bulk salt concentration $n_{b}$, such that the only variable is $\phi_{\pm}\left(0\right)\equiv \phi\left(0\right)$, the fraction of occupied sites at zero electrostatic potential (which is the same for positive and negative sites).

From the symmetry of the macro-ions, $Q\left(0\right)=0$. Substituting this  condition in Eqs.~(\ref{screening}) and~(\ref{rho2S}), it follows that the contribution of the macro-ions to the screening length is
\begin{equation}
\lambda_{\rm{p}}^{-2}=\left. -\frac{p_{b}}{\varepsilon_0 \varepsilon}\frac{Q\left(\psi\right)}{\partial\psi}\right|_{\psi=0}.
\label{lambdaMI}
\end{equation}
In order to find $Q(\psi)$ in terms of $\phi(0)$, we first invert the relation followed by Eqs.~(\ref{phi2S}) and (\ref{Z2S}),
\begin{equation}
\phi \left(0\right)=\frac{n_{b}K}{1+n_{b}K},
\end{equation}
and obtain
\begin{equation}
\label{alpha}
n_bK=\frac{\phi(0)}{1-\phi(0)} \, .
\end{equation}
Substituting Eq.~(\ref{alpha}) in Eq.~(\ref{ep2S}), we get
\begin{eqnarray}
Q\left(\psi\right) &=&Q_{+}(\psi)+ {Q_{-}(\psi)} \nonumber\\
& =& \frac{Ne\phi\left(0\right){\rm e}^{-\beta e\psi}}{1-\phi\left(0\right)\left(1-{\rm e}^{-\beta e\psi}\right)} -
\frac{Ne\phi\left(0\right){\rm e}^{\beta e\psi}}{1-\phi\left(0\right)\left(1-{\rm e}^{\beta e\psi}\right)}, \nonumber\\
\end{eqnarray}
and Eq.~(\ref{lambdaMI}) becomes
\begin{equation}
\lambda^{-2}_{\rm{p}}=\frac{2 p_b}{\varepsilon_0 \varepsilon}N\phi\left(0\right)\left(1-\phi\left(0\right)\right).
\label{variance}
\end{equation}

Equation~(\ref{variance}) shows that when the macro-ions are fully symmetric, their contribution to the screening length $\lambda_{\rm p}$ is proportional to the variance of the binomial distribution, $N\phi\left(0\right)\left(1-\phi\left(0\right)\right)$. The variance of this positive/negative charge distribution is maximal when $\phi\left(0\right)=1/2$ and minimal when $\phi\left(0\right)=0$ (or 1). Indeed, macro-ions that are ``half full" are free to adsorb or release charges, whereas completely ``full" or completely ``empty" macro-ions have less freedom; they can only release ions or only adsorb them, which reduces their ability to screen electrostatic interactions. This property explains the decrease followed by an increase of the screening length in Fig.~\ref{Fig5} (blue dashed curve).

Furthermore, if we take $k$ symmetric ions, each with occupancy fraction of $\phi(0)$, their screening is larger (smaller $\lambda_{\rm p}$) than for a single macro-ion with an occupancy fraction of $k\phi(0)$, because the relation
$k\phi\left(0\right)\left(1-\phi\left(0\right)\right)>k\phi\left(0\right)\left(1-k\phi\left(0\right)\right)$
~holds~ for ~any $k>1$. This property is responsible for the non-monotonic behavior of symmetric macro-ions that is shown in Fig.~\ref{Fig3} (solid black curve).

We finally remark that the above calculation is done for a fixed bulk density of ions, and the effect of the macro-ions on the bathing solution was not considered.


%
%
%
%

\end{document}